\documentclass[superscriptaddress,twocolumn,prl]{revtex4-2}
\usepackage{graphicx}
\usepackage{amsmath}
\usepackage{amssymb}
\usepackage{amsthm}
\usepackage[T1]{fontenc}
\usepackage[utf8]{inputenc}
\usepackage{lmodern}

\usepackage{hyperref}
 \hypersetup{
	colorlinks=true,
	linkcolor=red,
	filecolor=magenta,      
	urlcolor=magenta,
	citecolor=blue,
	breaklinks = true,
	backref = true,
	linktocpage=true   
}
\usepackage[all]{hypcap}
\usepackage{xcolor}

\begin{document}

%---------------------------------------------------------------------------------------------------------

\title {How to cleave cubic perovskite oxides}
%---------------------------------------------------------------------------------------------------------
\author{Igor Sokolović}
\email{igor.sokolovic@unileoben.ac.at}
\affiliation{Institute of Applied Physics, TU Wien, Wiedner Hauptstrasse 8-10/134, 1040 Vienna, Austria}

\author{Michael Schmid}
\affiliation{Institute of Applied Physics, TU Wien, Wiedner Hauptstrasse 8-10/134, 1040 Vienna, Austria}

\author{Ulrike Diebold}
\affiliation{Institute of Applied Physics, TU Wien, Wiedner Hauptstrasse 8-10/134, 1040 Vienna, Austria}

\author{Martin Setvín}
\affiliation{Institute of Applied Physics, TU Wien, Wiedner Hauptstrasse 8-10/134, 1040 Vienna, Austria}
\affiliation{Department of Surface and Plasma Science, Faculty of Mathematics and Physics, Charles University, 180 00 Prague 8, Czech Republic}

%---------------------------------------------------------------------------------------------------------
\begin{abstract} 
Surfaces of cubic perovskite oxides attract significant attention for their physical tunability and high potential for technical applications. Bulk-terminated surfaces are desirable for theoretical modelling and experimental reproducibility, yet there is a lack of methods for preparing such well-defined surfaces. We discuss a method for strain-assisted cleaving of perovskite single crystals, using a setup easily transferable between different experimental systems. The details of the cleaving device and the procedure were optimized in a systematic study on the model SrTiO$_3$. The large-area morphology and typical distribution of surface terminations on cleaved SrTiO$_3$(001) is presented, with specific guidelines on how to distinguish well-cleaved surfaces from conchoidally fractured ones. The cleaving is applicable to other cubic perovskites, as demonstrated on KTaO$_3$(001) and BaTiO$_3$(001). This approach opens up a pathway for obtaining high-quality surfaces of this promising class of materials.
\end{abstract}

\maketitle

%-------------------------------------------------------------------------
\begin{figure*} [t] 
	\begin{center} 
		\includegraphics[width=2.0\columnwidth,clip=false]{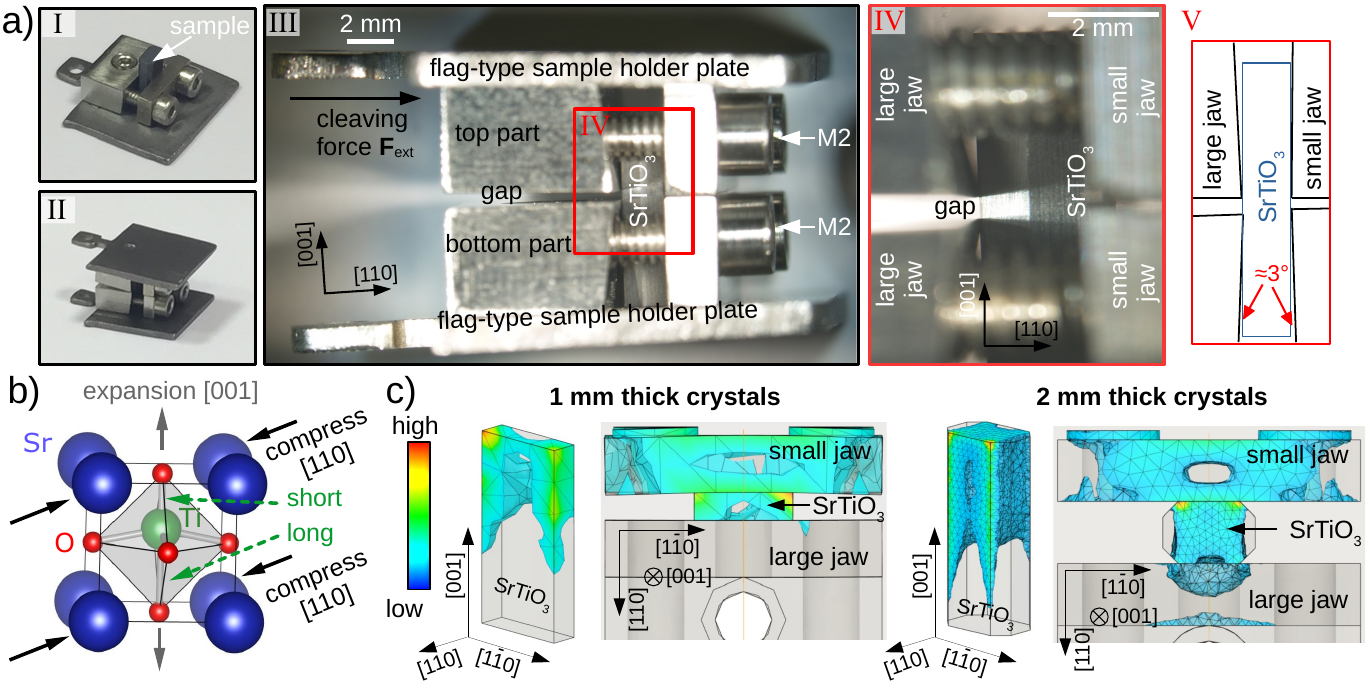}
	\end{center}
	\caption{a) Concept of the cleaving device. (I) SrTiO$_3$ crystal clamped at its bottom part; the device is mounted on a flag-type sample holder. (II) The upper part of the crystal is also strained by a mirror-symmetric clamp. (III) Side view on the cleaving device. (IV) Detailed view of the region indicated in ``III''. (V) Sketch of the same region.  b) Model showing the impact of the applied strain on the perovskite lattice (see text for details). c) Simulations of the static strain distribution after pre-straining a SrTiO$_3$ single crystal with sharp $\langle 001\rangle$ edges (left) and beveled edges (right). The color scale represents the magnitude of strain. Regions with strain below certain threshold are rendered transparent to highlight the points of interest. These transparent low-strain regions are strained rather homogeneously.} \label{fig:1} 
\end{figure*}
%------------------------------------------------------------------------- 

%-------------------------------------------------------------------------
\begin{figure*} [t] 
	\begin{center} 
		\includegraphics[width=2.0\columnwidth,clip=false]{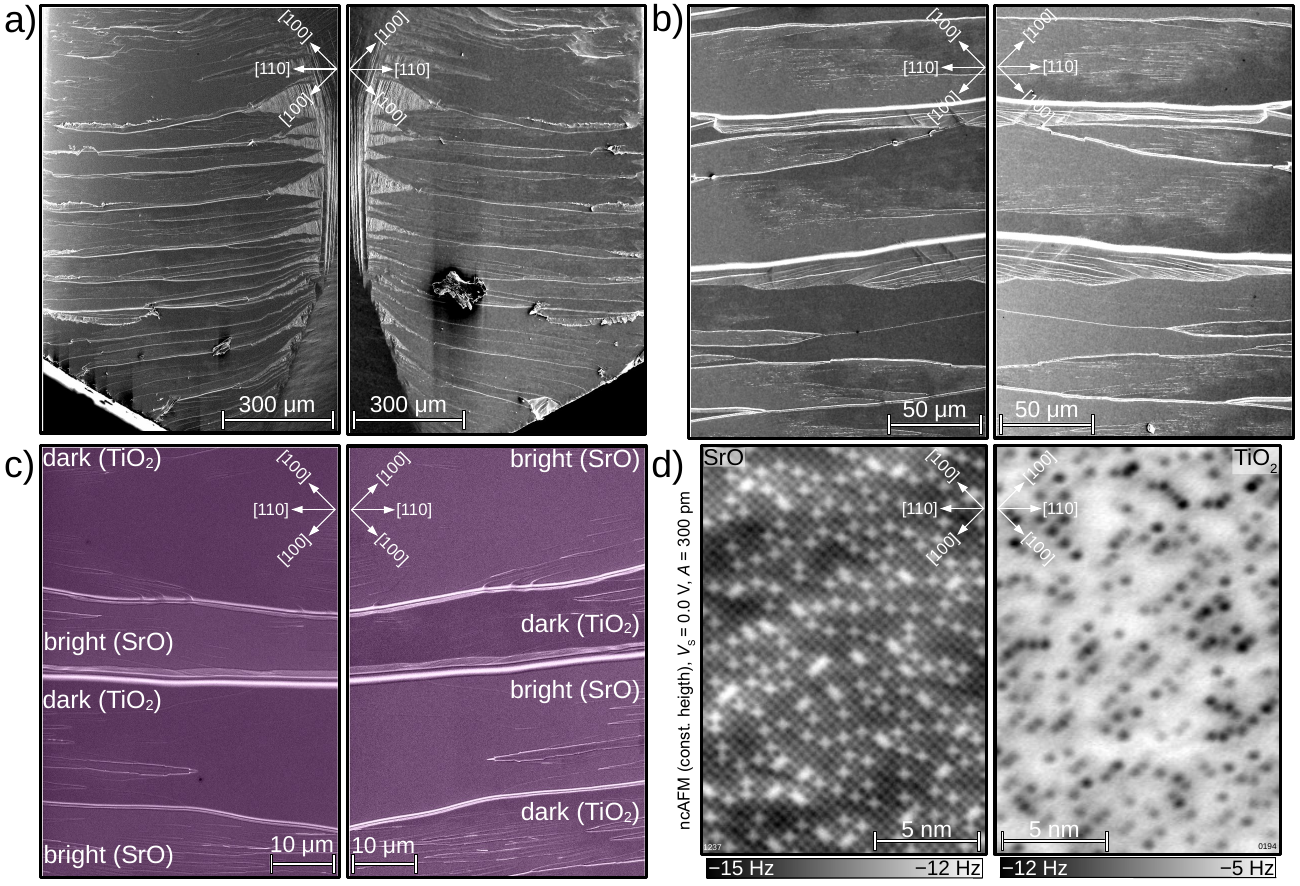}
	\end{center}
	\caption{Cleaving SrTiO$_3$(001) produces two mirror-symmetric surfaces. a--c) Secondary electron microscopy (SEM) images obtained on both sides of the same cleaved  SrTiO$_3$(001) surface. The morphology is mirrored and the distribution of surface terminations is the opposite on the two surfaces. The highest-magnification SEM image in (c) is colored for better visibility. d) Atomically resolved ncAFM images of both surface terminations.} \label{fig:2}
\end{figure*}
%------------------------------------------------------------------------- 

%-------------------------------------------------------------------------
\begin{figure*} [t] 
	\begin{center} 
		\includegraphics[width=2.0\columnwidth,clip=false]{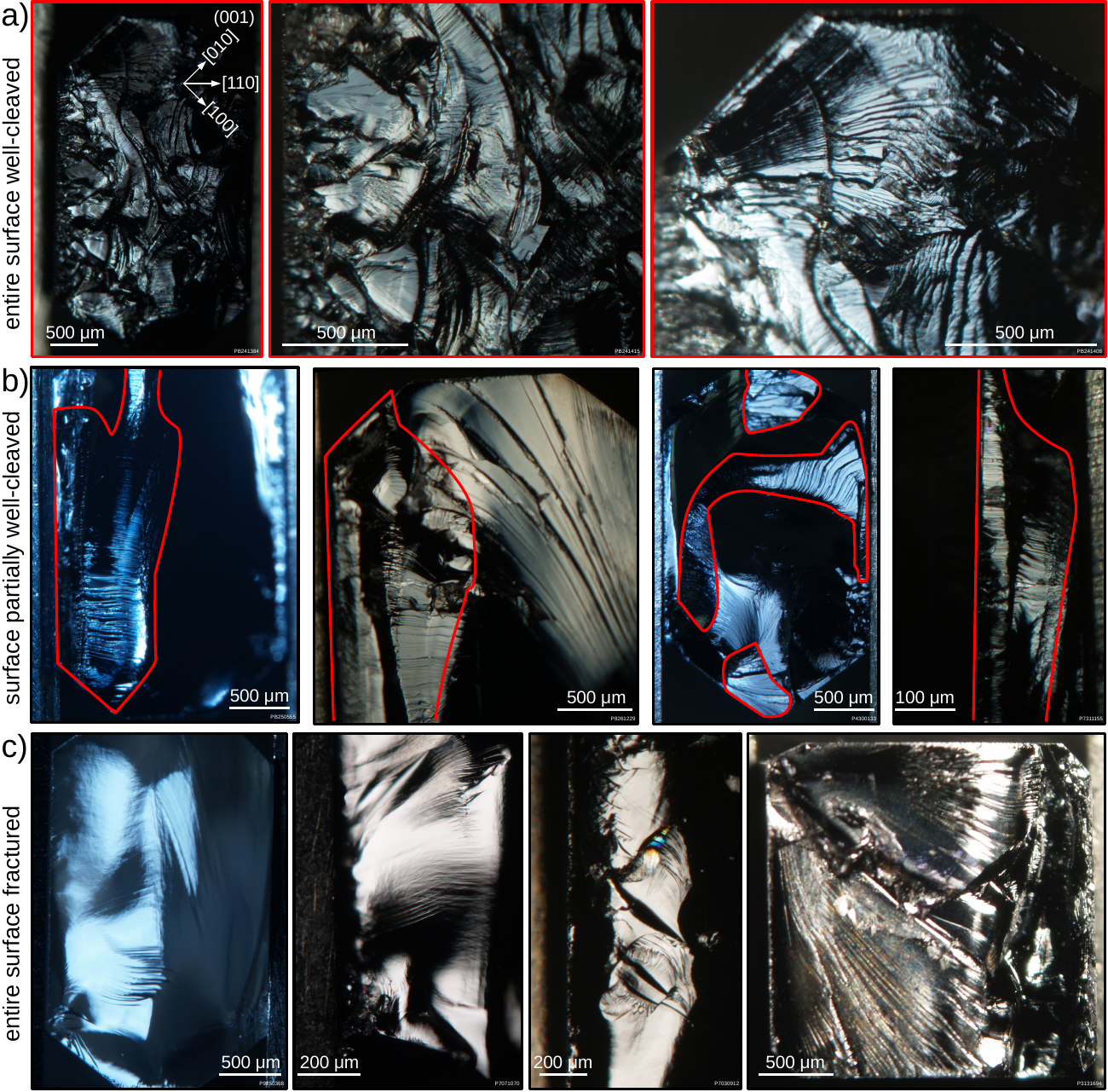}
	\end{center}
	\caption{Distinction between well-cleaved and fractured SrTiO$_3$(001) surfaces on optical photographs. Cleaved SrTiO$_3$(001) surfaces categorized according to the quality of the cleave: a) well-cleaved surface regions span the majority of the surface (first panel shows the whole surface and the remaining two focus on the details), b) partially cleaved and partially fractured surface (well-cleaved areas are outlined in red), c) completely fractured surface.} \label{fig:3}
\end{figure*}
%------------------------------------------------------------------------- 

%-------------------------------------------------------------------------
\begin{figure*} [t] 
	\begin{center} 
		\includegraphics[width=2.0\columnwidth,clip=false]{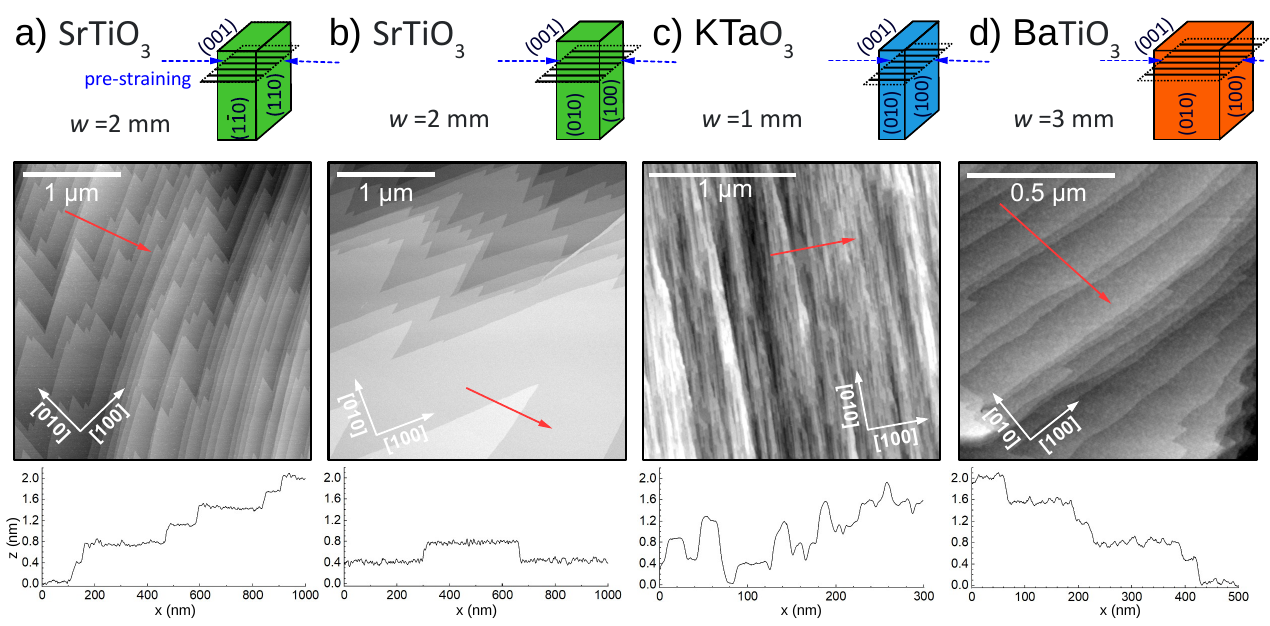}
	\end{center}
	\caption{Ambient tapping mode AFM images of cleaved (001) surfaces of three perovskite oxides: a,b) SrTiO$_3$ (two orientations), c) KTaO$_3$, d) BaTiO$_3$. The geometry and the width $w$ of the cleaved single crystals is sketched in the top row; the direction of strain is marked by blue arrows. Line profiles along the red arrows marked in the AFM images are plotted in the bottom row. Note that these large-area images are affected by distortions due to creep of the piezoelectric scanner in the x-y plane; the z-creep has been corrected.} 
	\label{fig:4}
\end{figure*}
%------------------------------------------------------------------------- 

Cleaving is a preferred preparation method for many surfaces, because cleaving $in$ $situ$, $i.e.$ in ultrahigh vacuum (UHV) creates perfectly clean surfaces that do not require further treatment \cite{lander1963structural,haneman1968electron, halwidl2017ordered, dulub2010preparation, barivsic2008demonstrating,barivsic2013universal,yamada2015dynamic,yan2019crystal}. A wide range of solids with preferable cleavage planes readily cleave \cite{gandhi1983fracture}: two-dimensional materials such as graphite and graphene cleave/peal \cite{hernandez2008high} by breaking the inter-layer van der Waals forces. Layered materials like muscovite mica \cite{ostendorf2008flat,franceschi2023resolving} or any member of the Ruddlesden–Popper perovskite series \cite{halwidl2016adsorption} cleave along the layer-separating planes. Ionic crystals like NaCl \cite{ino1964epitaxial} cleave $via$ bond cleavage along the main crystallographic planes. These materials can be cleaved using a sharp blade or can even be exfoliated. On the other hand, materials that do not posses a natural cleavage plane tend to  break by the conchoidal fracture, $i.e.$, in direction that does not follow any crystallographic plane and the resulting surfaces lack well-defined structure. Cubic perovskite oxides, such as SrTiO$_3$, are a typical example. Since cleaving these materials is not straightforward, there is a systematic absence of their well-defined bulk-truncated surfaces in literature, despite the vast interest in their use in oxide electronics \cite{takagi2010emergent,noel2020non,chen2015extreme,varignon2018new}, catalysis \cite{wang2021perovskite} or solid oxide fuel cells \cite{jun2016perovskite,shu2019advanced}. Here, we present a procedure that efficiently cleaves cubic perovskite oxides, and that is readily transferable between different experimental measurement systems. The principle of operation is discussed and demonstrated on strontium titanate SrTiO$_3$, and successful application to KTaO$_3$ and BaTiO$_3$ promises universality.

Cubic perovskite oxides with the ABO$_3$ stoichiometry posses inversion symmetry and no natural cleavage planes. Moreover, they are commonly characterized by mixed ionic and covalent bonding within each unit cell: while A--O interaction is electrostatic, bonds in BO$_6$ octahedra have a highly covalent character. The model cubic perovskite oxide SrTiO$_3$ conchoidally fractures when supplied with a strong-enough external force, yielding surfaces that are ill-defined at the atomic level. Recently, we managed to cleave SrTiO$_3$ single crystals and create atomically flat (001) surfaces \cite{sokolovic2019incipient,sokolovic2021quest} by engineering strain in the cleavage region. An optimized setup for cleaving samples in ultra-high vacuum (UHV) is described in Fig. \ref{fig:1}. The SrTiO$_3$ single crystal is pre-strained prior to insertion into the vacuum chamber, see the photos in Fig.\,\hyperref[fig:1]{\ref{fig:1}a}. The cleaving device consists of two identical clamps, one of which compresses the sample at the bottom (photo ``I'') and the other one at the top (photo ``II''). Each clamp consists of a larger jaw that is fixed to its respective flag-type sample holder plate, and a smaller jaw that compresses the single crystal by tightening the screws connecting the small and the large jaws. The larger jaws are polished under an $\approx 5^\circ$ angle at the face that is oriented towards a single crystal, in order to minimize the contact area with the jaw and an increase the contact pressure (see ``IV'' and ``V'' in Fig.\,\hyperref[fig:1]{\ref{fig:1}a}). The single crystal needs to be subjected to sufficient pressure during the pre-straining procedure: a good indicator is the tilting of the sample by $\approx 2-5^\circ$ from the surface normal of the bottom sample plate (``V'' in Fig.\,\hyperref[fig:1]{\ref{fig:1}a}) to accommodate for the geometry of the large jaws. The jaws can be fabricated from any material. Austenitic stainless steel provided better results than molybdenum or various aluminum alloys, presumably due to its suitable mechanical properties. 

A proposed atomic-scale model of the involved mechanisms is outlined in Fig.\,\hyperref[fig:1]{\ref{fig:1}b}. The strain applied by the clamps compresses the lattice and facilitates its expansion in the perpendicular direction (transverse expansion, Poisson's ratio). In principle, this can weaken the corresponding bonds along the c axis and support cleavage in the region where the sample is compressed. This mechanism may be in principle valid in any material, while perovskites possess properties that further promote creation of a cleavage plane: At sufficiently high strain, expansion of the lattice constant along the $c$ axis can induce ferroelectric polarization \cite{muller1968characteristic,muller1979srtio3,burke1971stress, haeni2004room, griffith1921phenomena}, where the transition metal atom in the centre of the cubic cell (Ti in Fig.\,\hyperref[fig:1]{\ref{fig:1}b}) shifts in direction parallel to the lattice expansion. This leads to a further elongation of one of the covalent bonds in the Ti octahedron and breaking the cubic crystal symmetry. We note that the strain provided solely by the clamps is not sufficient to drive SrTiO$_3$ to the ferroelectric phase at room temperature; the force applicable by the used M2 screws is an order of magnitude smaller \cite{Meza2024, haeni2004room, griffith1921phenomena, sokolovic2019incipient}. The ferroelectric polarization can only be driven by the tensile strain present at the tip of the crack that propagates through the crystal. The strain applied by the clamps only plays a supportive role and confines the crack propagation in a well-defined gap between the top and the bottom clamps  (``IV'' in Fig.\,\hyperref[fig:1]{\ref{fig:1}a}). The pre-straining procedure is laid out thoroughly in the Supporting Information (including Methods).

The applied strains are large enough to induce plastic deformation of the steel jaws. The plastic deformation is typically visible as imprints in the steel jaws and they must be repolished after each use. Figure\,\hyperref[fig:1]{\ref{fig:1}c} provides an impression of the strain distribution in the SrTiO$_3$ crystal. The highest strain appears at the edges of the SrTiO$_3$ crystal, see the left panel in Fig. \,\hyperref[fig:1]{\ref{fig:1}c}. To avoid premature (and ill-defined) cracking at these high-strain points, we use samples with beveled edges parallel to $\langle$001$\rangle$, which result in a more homogeneous strain distribution, see the right panel in Fig.\,\hyperref[fig:1]{\ref{fig:1}c}. Using this procedure, we routinely cleave 2\,mm thick samples (all surfaces shown in this paper) to create large surface areas, desirable for area-averaging techniques such as x-ray photoelectron spectroscopy or electron diffraction. Thin crystals (below 1\,mm) are generally easier to cleave, even if the sharp edges are kept. 

Each cleavage creates two mirror-symmetric surfaces, and the double-decker construction with two flag-type sample holders (Fig.\,\hyperref[fig:1]{\ref{fig:1}a}) enables utilization of both sides of the cleaved crystals. Figure\,\ref{fig:2} shows scanning electron microscopy (SEM) images of both surfaces side by side. The morphology of the two surfaces is mirrored, and they exhibit an opposite SEM contrast.  Cleaving SrTiO$_3$ along the (001) plane always results in one TiO$_2$- and one SrO-terminated side, as these are the only possible surface terminations. This leads to symmetry breaking along the [001] direction. Hence, the SEM contrast differences stem from the opposite distribution of surface terminations on the two sides of a same cleaved surface. Namely, if a surface region on one side is TiO$_2$-terminated, the same region on the other side of a cleaved SrTiO$_3$(001) surface must be terminated with a SrO layer. Although chemically differing in just one atomic layer, the two terminations are recognizable with SEM due to their different  work functions and electronic structure: TiO$_2$ is metallic and SrO semiconducting upon cleavage in UHV \cite{sokolovic2019incipient}. For obtaining the SEM images, the samples had to be transported through air, however, which is likely to modify the surfaces; nevertheless they show a clearly different brightness in SEM, even after exposing them to ambient conditions for one month.

The SEM images in Figs.\,\hyperref[fig:2]{\ref{fig:2}a--c} also demonstrate that our SrTiO$_3$ cleaving procedure creates heterogeneous (001) surfaces. They consist of macroscopically large, alternating domains of metallic and semiconducting material, providing a versatile substrate for application in electronics or catalysis. In fact, the two terminations likely preferentially adsorb different atmospheric gases, as they can still be recognized even after the cleaved SrTiO$_3$(001) surface was exposed to ambient for a prolonged period of time (months) prior to the measurements in Fig.\,\ref{fig:2}. 

The well-cleaved SrTiO$_3$(001) surface areas $a$ $priori$ have complementary atomic structures: SrO and TiO$_2$ terminations are always covered by $0.14\pm0.02$ monolayers of Sr vacancies and adatoms, respectively \cite{sokolovic2019incipient}. We have argued that they compensate for the surface polarity during cleaving \cite{sokolovic2019incipient}. 
Figure\,\hyperref[fig:2]{\ref{fig:2}d} shows atomically resolved non-contact atomic force microscopy (ncAFM) \cite{setvin2012ultrasharp, Giessibl2019Review, huber2017low, schmid2019device} images of the two terminations, obtained with an O-terminated tip \cite{sokolovic2020resolving}, which detects Sr atoms as attractive (dark), O atoms as repulsive (bright), and missing Sr atoms as a lack of attractive interaction (bright ``x''-like features). We find that this atomic-scale surface structure is insensitive to details of the preparation. Up to now, we have cleaved more than 40 SrTiO$_3$ single crystals with 0.5.\,wt\% of Nb doping, obtained from various vendors (Supplement). The quality of single crystals can affect the quality of cleaved SrTiO$_3$(001) surfaces only in the relative coverage of well-cleaved areas with respect to conchoidally fractured surface areas. However, the atomic structure of the well-cleaved surface areas was always the same.

The crack front of a cleavage cannot be controlled during its propagation. It is common that a cleaved surface contains both, areas with conchoidal fracture and well-cleaved surface; this holds especially true for thick samples where the pre-straining is less efficient. No two cleavages are ever the same, although almost all ($>90$\% of cleaved surfaces) contain at least some well-cleaved regions. The cleaved and fractured portions of a SrTiO$_3$(001) surface can be readily distinguished with an optical microscope, as shown in Fig.\,\ref{fig:3}. The identification is even easier under a stereo microscope when the angle of illumination can be varied: samples in Fig.\,\ref{fig:3} are illuminated under a specific angle where the light is reflected from the flat (001) terraces, highlighting the well-cleaved regions.
The best SrTiO$_3$(001) cleavages are either macroscopically flat (as the one in \cite{sokolovic2019incipient}), or contain many separated, well-cleaved regions of step bunching, as shown  Fig.\,\hyperref[fig:3]{\ref{fig:3}a}. A train of terraces separated by optically-visible straight, or partially straight steps parallel to each other is a fingerprint of a well-cleaved region (almost all bright regions in Fig.\,\hyperref[fig:3]{\ref{fig:3}a} and outlined regions in Fig.\,\hyperref[fig:3]{\ref{fig:3}b}). They commonly span laterally for hundreds of micrometers, or several millimetres (Fig.\,\hyperref[fig:2]{\ref{fig:2}a}). Most of these areas have steps running parallel to the force used for pre-straining the crystal (roughly the $[110]$ direction in Fig.\,\ref{fig:3}), but even if not, the steps are always mutually parallel on a well-cleaved surface patch. In contrast, fractured surface regions commonly appear smooth and featureless at a macroscopic scale. The ill-defined surface patches (other than areas highlighted in red in Fig.\,\ref{fig:3}) appear in many flavours, with inconsistent atomic structure and electronic properties.

Successive criteria for distinguishing good from bad cleaved surface areas can be inferred from the systematic overview in Fig.\,\ref{fig:3}: (\textit{I}) well-cleaved areas reflect light best when illuminated and viewed from the top; when the illumination is not fixed to a [001] axis, all well-cleaved (001) surface regions reflect light simultaneously at certain angles. (\textit{II}) Well-cleaved areas are not smooth but have optically visible steps separating atomically flat terraces. (\textit{III}) The steps on well-cleaved areas are parallel to each other and commonly form a terrace train. (\textit{IV}) If the fractured surface areas are not smooth but have large optically visible steps instead, conchoidal fracture can be recognized by the reflection of light under many different angles due to the random inclination with respect to the [001] surface normal. 

The double-decker construction of a cleaving device (Fig.\,\ref{fig:1}) allows removal of one half of the double-decker from a UHV chamber and its characterization with various techniques to provide better guidance for studying the mirror-symmetric counterpart that remains in the UHV chamber. This is a highly desirable option, since the quality of a cleavage and the distribution of well-cleaved areas cannot be predicted $a$ $priori$ (Fig.\,\ref{fig:3}). It ensures that the UHV measurements are performed on well-defined SrTiO$_3$(001) surfaces instead on the randomly inclined regions of conchiodal fracture, ill-defined at the atomic level, which are unavoidable on any cleaved SrTiO$_3$(001).

Previous methods for cleaving SrTiO$_3$(001) \cite{chien2009controllable,chien2010survey,guisinger2009nanometer,sitaputra2015topographic,hunter2024controlling} were based on hitting a thin ($<250$\,$\mu$m thick) SrTiO$_3$ single crystal in UHV  strongly enough that it breaks, preferably at temperatures below the  cubic-tetragonal phase transition at 105\,K \cite{sakudo1971dielectric,aschauer2014competition}. The double-decker cleaving device (Fig.\,\ref{fig:1}) provides significant improvements to this method. First, cleaving is readily performed at room temperature. Next, thick crystals can be cleaved to yield large (>1\,mm$^2$) well-cleaved surface areas suitable for area-averaging methods. Finally, the double-decker construction enables removing the mirror-symmetric counter-piece from the UHV, providing the possibility to assess the quality of the cleave, and directing the measurements towards properly cleaved surface areas, or performing complementary measurements by different experimental techniques.

Another advantage of our procedure is that a bulk-truncation of SrTiO$_3$ along the [001] direction is guaranteed in the well-cleaved regions. It overcomes the pitfalls of wet-chemical etching of the cut-and-polished SrTiO$_3$(001) substrates \cite{kawasaki1994atomic,connell2012preparation,deak2006ordering} that can lead to unintentional contamination \cite{chambers2012unintentional}. Our approach avoids the occurrence of many surface reconstructions caused by the cleaning procedures (sputtering, annealing) necessary after introducing samples into UHV \cite{castell2002scanning,silly2006srtio3,erdman2002structure,herger2007surfacePRL, Gerhold2014, Goniakowski2008}. In fact, we recently showed that the ($1\times 1$) termination is lost on such surfaces, even without sputtering, once the contaminants are thermally removed in UHV, and that the ($1\times 1$) diffraction pattern in LEED is not indicative of the unreconstructed surface because it can also originate from subsurface layers \cite{sokolovic2021quest}. Chemical etching removes one of the two terminations \cite{koster1998influence,koster1998quasi,baniecki2008photoemission,castell2002scanning}, while cleaving produces clean, well-defined SrTiO$_3$(001) surfaces with both, TiO$_2$- and SrO- terminated regions (Fig.\,\ref{fig:2}), a heterogeneous substrate suitable for physical and catalytic characterization \cite{sokolovic2024duality}.

The principles behind cleaving SrTiO$_3$ with the cleaving device in Fig.\,\ref{fig:1} seem intrinsic to all (pseudo)cubic perovskite oxides, due to the common flexibility of the B cations in their lattices in response to strain \cite{cohen1992origin,resta1993towards,bousquet2008improper,hwang2019tuning}. 
The AFM images (taken under ambient conditions) in Fig.\,\ref{fig:4} show cleaved (001) surfaces of SrTiO$_3$, BaTiO$_3$, and KTaO$_3$ achieved through the use of the cleaving device in Fig.\,\ref{fig:1}. All are cubic perovskites, with SrTiO$_3$ and BaTiO$_3$ nominally non-polar along the [001] direction (when neglecting the ferroelectric distortions of BaTiO$_3$), and KTaO$_3$, where unreconstructed (001) surfaces are polar \cite{uwe1973stress,vendik1998microwave,setvin2018polarity} and one of the mechanisms to compensate the polarity is a high step density. 
All materials exposed atomically flat surface areas with unit-cell high steps ($\approx$0.4 nm) when cleaved this way. Fig.\,\hyperref[fig:4]{\ref{fig:4}a} shows a result of cleaving a SrTiO$_3$(001) single crystal in the geometry used throughout this manuscript, where the compressive strain was applied parallel to the [110] direction. The image was taken on one of the optically visible flat terraces such are those in Figs.\,\ref{fig:2}\,and\,\ref{fig:3}. SrTiO$_3$ single crystals were also cleaved in a different geometry, where the strain was applied parallel to the [100] direction to a single crystal cut in a way to expose \{001\} surfaces (Fig.\,\hyperref[fig:4]{\ref{fig:4}b}). Cleaved KTaO$_3$(001) developed narrow terraces, while the atomically flat terraces are significantly larger on cleaved BaTiO$_3$(001). In both cases the step orientation is approximately parallel to the application of strain, along the [100] direction in this case. It is noteworthy that KTaO$_3$(001), unlike SrTiO$_3$, can also develop atomically flat surfaces after cleaving with a sharp tungsten carbide blade \cite{setvin2018polarity}, which could hold true for the truly ferroelectric BaTiO$_3$(001) as well. In either case, the application of the cleaving device still provides advantages with respect to its transferability and the creation of two research-ready surfaces.

In conclusion, the procedure described in this work routinely produces bulk-terminated (001) surfaces of perovskites ABO$_3$ with large, atomically flat surface regions of both, AO and BO$_2$ terminations. Well-cleaved SrTiO$_3$(001) surface areas were found to always have the same chemical and physical properties, granting $a$ $priori$ knowledge of the surface structure. The distribution of the surface termination on cleaved SrTiO$_3$ can be observed by SEM, while the conchoidally fractured and well-cleaved areas can be distinguished by an optical microscope alone. The creation of two mirror-symmetric, research-ready surfaces with each cleaving allows simultaneous measurements on two surfaces with opposite terminations. The method of cleaving by engineering the crack propagation $via$ pre-straining single crystals was successfully applied to three cubic perovskite oxides SrTiO$_3$, KTaO$_3$ and BaTiO$_3$, suggesting general applicability. Together with the transferability of the cleaving device, this provides a convenient and reproducible pathway to obtaining truly bulk-terminated surfaces of this wide family of solids.
 
This research was funded in part by the Austrian Science Fund (FWF) projects SuPer (P32148-N36) and SFB Project TACO (Grant-DOI:10.55776/F81). For open access purposes, the author has applied a CC BY public copyright license to any author accepted manuscript version arising from this submission. M.Se. acknowledges support from the Czech Science Foundation GACR 20-21727X and MSMT OP JAK CZ$.02.01.01/00/22\_008/0004572$. The authors acknowledge the expertise of Herbert Schmid, Reiner Gärtner and Martin Macharacek in fabricating cleaving devices from various materials.

\bibliography{Bib_HowToCleaveCubicPerovskiteOxides.bib}

\begin{thebibliography}{62}%
\makeatletter
\providecommand \@ifxundefined [1]{%
 \@ifx{#1\undefined}
}%
\providecommand \@ifnum [1]{%
 \ifnum #1\expandafter \@firstoftwo
 \else \expandafter \@secondoftwo
 \fi
}%
\providecommand \@ifx [1]{%
 \ifx #1\expandafter \@firstoftwo
 \else \expandafter \@secondoftwo
 \fi
}%
\providecommand \natexlab [1]{#1}%
\providecommand \enquote  [1]{``#1''}%
\providecommand \bibnamefont  [1]{#1}%
\providecommand \bibfnamefont [1]{#1}%
\providecommand \citenamefont [1]{#1}%
\providecommand \href@noop [0]{\@secondoftwo}%
\providecommand \href [0]{\begingroup \@sanitize@url \@href}%
\providecommand \@href[1]{\@@startlink{#1}\@@href}%
\providecommand \@@href[1]{\endgroup#1\@@endlink}%
\providecommand \@sanitize@url [0]{\catcode `\\12\catcode `\$12\catcode
  `\&12\catcode `\#12\catcode `\^12\catcode `\_12\catcode `\%12\relax}%
\providecommand \@@startlink[1]{}%
\providecommand \@@endlink[0]{}%
\providecommand \url  [0]{\begingroup\@sanitize@url \@url }%
\providecommand \@url [1]{\endgroup\@href {#1}{\urlprefix }}%
\providecommand \urlprefix  [0]{URL }%
\providecommand \Eprint [0]{\href }%
\providecommand \doibase [0]{https://doi.org/}%
\providecommand \selectlanguage [0]{\@gobble}%
\providecommand \bibinfo  [0]{\@secondoftwo}%
\providecommand \bibfield  [0]{\@secondoftwo}%
\providecommand \translation [1]{[#1]}%
\providecommand \BibitemOpen [0]{}%
\providecommand \bibitemStop [0]{}%
\providecommand \bibitemNoStop [0]{.\EOS\space}%
\providecommand \EOS [0]{\spacefactor3000\relax}%
\providecommand \BibitemShut  [1]{\csname bibitem#1\endcsname}%
\let\auto@bib@innerbib\@empty
%</preamble>
\bibitem [{\citenamefont {Lander}\ \emph {et~al.}(1963)\citenamefont {Lander},
  \citenamefont {Gobeli},\ and\ \citenamefont
  {Morrison}}]{lander1963structural}%
  \BibitemOpen
  \bibfield  {author} {\bibinfo {author} {\bibfnamefont {J.~J.}\ \bibnamefont
  {Lander}}, \bibinfo {author} {\bibfnamefont {G.~W.}\ \bibnamefont {Gobeli}},\
  and\ \bibinfo {author} {\bibfnamefont {J.}~\bibnamefont {Morrison}},\
  }\bibfield  {title} {\bibinfo {title} {Structural properties of cleaved
  silicon and germanium surfaces},\ }\href@noop {} {\bibfield  {journal}
  {\bibinfo  {journal} {J. Appl. Phys.}\ }\textbf {\bibinfo {volume} {34}},\
  \bibinfo {pages} {2298} (\bibinfo {year} {1963})}\BibitemShut {NoStop}%
\bibitem [{\citenamefont {Haneman}(1968)}]{haneman1968electron}%
  \BibitemOpen
  \bibfield  {author} {\bibinfo {author} {\bibfnamefont {D.}~\bibnamefont
  {Haneman}},\ }\bibfield  {title} {\bibinfo {title} {Electron paramagnetic
  resonance from clean single-crystal cleavage surfaces of silicon},\
  }\href@noop {} {\bibfield  {journal} {\bibinfo  {journal} {Phys. Rev.}\
  }\textbf {\bibinfo {volume} {170}},\ \bibinfo {pages} {705} (\bibinfo {year}
  {1968})}\BibitemShut {NoStop}%
\bibitem [{\citenamefont {Halwidl}\ \emph {et~al.}(2017)\citenamefont
  {Halwidl}, \citenamefont {Mayr-Schm{\"o}lzer}, \citenamefont {Fobes},
  \citenamefont {Peng}, \citenamefont {Mao}, \citenamefont {Schmid},
  \citenamefont {Mittendorfer}, \citenamefont {Redinger},\ and\ \citenamefont
  {Diebold}}]{halwidl2017ordered}%
  \BibitemOpen
  \bibfield  {author} {\bibinfo {author} {\bibfnamefont {D.}~\bibnamefont
  {Halwidl}}, \bibinfo {author} {\bibfnamefont {W.}~\bibnamefont
  {Mayr-Schm{\"o}lzer}}, \bibinfo {author} {\bibfnamefont {D.}~\bibnamefont
  {Fobes}}, \bibinfo {author} {\bibfnamefont {J.}~\bibnamefont {Peng}},
  \bibinfo {author} {\bibfnamefont {Z.}~\bibnamefont {Mao}}, \bibinfo {author}
  {\bibfnamefont {M.}~\bibnamefont {Schmid}}, \bibinfo {author} {\bibfnamefont
  {F.}~\bibnamefont {Mittendorfer}}, \bibinfo {author} {\bibfnamefont
  {J.}~\bibnamefont {Redinger}},\ and\ \bibinfo {author} {\bibfnamefont
  {U.}~\bibnamefont {Diebold}},\ }\bibfield  {title} {\bibinfo {title} {Ordered
  hydroxyls on {Ca}$_3${Ru}$_2${O}$_7$(001)},\ }\href@noop {} {\bibfield
  {journal} {\bibinfo  {journal} {Nature Communications}\ }\textbf {\bibinfo
  {volume} {8}},\ \bibinfo {pages} {23} (\bibinfo {year} {2017})}\BibitemShut
  {NoStop}%
\bibitem [{\citenamefont {Dulub}\ and\ \citenamefont
  {Diebold}(2010)}]{dulub2010preparation}%
  \BibitemOpen
  \bibfield  {author} {\bibinfo {author} {\bibfnamefont {O.}~\bibnamefont
  {Dulub}}\ and\ \bibinfo {author} {\bibfnamefont {U.}~\bibnamefont
  {Diebold}},\ }\bibfield  {title} {\bibinfo {title} {Preparation of a pristine
  {TiO}$_2$ anatase (101) surface by cleaving},\ }\href@noop {} {\bibfield
  {journal} {\bibinfo  {journal} {J Phys. Cond. Mater.}\ }\textbf {\bibinfo
  {volume} {22}},\ \bibinfo {pages} {084014} (\bibinfo {year}
  {2010})}\BibitemShut {NoStop}%
\bibitem [{\citenamefont {Bari{\v{s}}i{\'c}}\ \emph {et~al.}(2008)\citenamefont
  {Bari{\v{s}}i{\'c}}, \citenamefont {Li}, \citenamefont {Zhao}, \citenamefont
  {Chabot-Couture}, \citenamefont {Yu},\ and\ \citenamefont
  {Greven}}]{barivsic2008demonstrating}%
  \BibitemOpen
  \bibfield  {author} {\bibinfo {author} {\bibfnamefont {N.}~\bibnamefont
  {Bari{\v{s}}i{\'c}}}, \bibinfo {author} {\bibfnamefont {Y.}~\bibnamefont
  {Li}}, \bibinfo {author} {\bibfnamefont {Y.-C.}\ \bibnamefont {Zhao},
  \bibfnamefont {X.and~Cho}}, \bibinfo {author} {\bibfnamefont
  {G.}~\bibnamefont {Chabot-Couture}}, \bibinfo {author} {\bibfnamefont
  {G.}~\bibnamefont {Yu}},\ and\ \bibinfo {author} {\bibfnamefont
  {M.}~\bibnamefont {Greven}},\ }\bibfield  {title} {\bibinfo {title}
  {Demonstrating the model nature of the high-temperature superconductor
  {H}g{B}a$_2${CuO}$_{4+ \delta}$},\ }\href@noop {} {\bibfield  {journal}
  {\bibinfo  {journal} {Phys. Rev. B}\ }\textbf {\bibinfo {volume} {78}},\
  \bibinfo {pages} {054518} (\bibinfo {year} {2008})}\BibitemShut {NoStop}%
\bibitem [{\citenamefont {Bari{\v{s}}i{\'c}}\ \emph {et~al.}(2013)\citenamefont
  {Bari{\v{s}}i{\'c}}, \citenamefont {Chan}, \citenamefont {Li}, \citenamefont
  {Yu}, \citenamefont {Zhao}, \citenamefont {Dressel}, \citenamefont
  {Smontara},\ and\ \citenamefont {Greven}}]{barivsic2013universal}%
  \BibitemOpen
  \bibfield  {author} {\bibinfo {author} {\bibfnamefont {N.}~\bibnamefont
  {Bari{\v{s}}i{\'c}}}, \bibinfo {author} {\bibfnamefont {M.~K.}\ \bibnamefont
  {Chan}}, \bibinfo {author} {\bibfnamefont {Y.}~\bibnamefont {Li}}, \bibinfo
  {author} {\bibfnamefont {G.}~\bibnamefont {Yu}}, \bibinfo {author}
  {\bibfnamefont {X.}~\bibnamefont {Zhao}}, \bibinfo {author} {\bibfnamefont
  {M.}~\bibnamefont {Dressel}}, \bibinfo {author} {\bibfnamefont
  {A.}~\bibnamefont {Smontara}},\ and\ \bibinfo {author} {\bibfnamefont
  {M.}~\bibnamefont {Greven}},\ }\bibfield  {title} {\bibinfo {title}
  {Universal sheet resistance and revised phase diagram of the cuprate
  high-temperature superconductors},\ }\href@noop {} {\bibfield  {journal}
  {\bibinfo  {journal} {PNAS}\ }\textbf {\bibinfo {volume} {110}},\ \bibinfo
  {pages} {12235} (\bibinfo {year} {2013})}\BibitemShut {NoStop}%
\bibitem [{\citenamefont {Yamada}\ \emph {et~al.}(2015)\citenamefont {Yamada},
  \citenamefont {Yamada}, \citenamefont {Phuong}, \citenamefont {Maruyama},
  \citenamefont {Nishimura}, \citenamefont {Wakamiya}, \citenamefont {Murata},\
  and\ \citenamefont {Kanemitsu}}]{yamada2015dynamic}%
  \BibitemOpen
  \bibfield  {author} {\bibinfo {author} {\bibfnamefont {Y.}~\bibnamefont
  {Yamada}}, \bibinfo {author} {\bibfnamefont {T.}~\bibnamefont {Yamada}},
  \bibinfo {author} {\bibfnamefont {L.~Q.}\ \bibnamefont {Phuong}}, \bibinfo
  {author} {\bibfnamefont {N.}~\bibnamefont {Maruyama}}, \bibinfo {author}
  {\bibfnamefont {H.}~\bibnamefont {Nishimura}}, \bibinfo {author}
  {\bibfnamefont {A.}~\bibnamefont {Wakamiya}}, \bibinfo {author}
  {\bibfnamefont {Y.}~\bibnamefont {Murata}},\ and\ \bibinfo {author}
  {\bibfnamefont {Y.}~\bibnamefont {Kanemitsu}},\ }\bibfield  {title} {\bibinfo
  {title} {Dynamic optical properties of {CH}$_3${NH}$_3${PbI}$_3$ single
  crystals as revealed by one-and two-photon excited photoluminescence
  measurements},\ }\href@noop {} {\bibfield  {journal} {\bibinfo  {journal}
  {JACS}\ }\textbf {\bibinfo {volume} {137}},\ \bibinfo {pages} {10456}
  (\bibinfo {year} {2015})}\BibitemShut {NoStop}%
\bibitem [{\citenamefont {Yan}\ \emph {et~al.}(2019)\citenamefont {Yan},
  \citenamefont {Zhang}, \citenamefont {Heitmann}, \citenamefont {Huang},
  \citenamefont {Chen}, \citenamefont {Cheng}, \citenamefont {Wu},
  \citenamefont {Vaknin}, \citenamefont {Sales},\ and\ \citenamefont
  {McQueeney}}]{yan2019crystal}%
  \BibitemOpen
  \bibfield  {author} {\bibinfo {author} {\bibfnamefont {J.-Q.}\ \bibnamefont
  {Yan}}, \bibinfo {author} {\bibfnamefont {Q.}~\bibnamefont {Zhang}}, \bibinfo
  {author} {\bibfnamefont {T.}~\bibnamefont {Heitmann}}, \bibinfo {author}
  {\bibfnamefont {Z.}~\bibnamefont {Huang}}, \bibinfo {author} {\bibfnamefont
  {K.~Y.}\ \bibnamefont {Chen}}, \bibinfo {author} {\bibfnamefont {J.-G.}\
  \bibnamefont {Cheng}}, \bibinfo {author} {\bibfnamefont {W.}~\bibnamefont
  {Wu}}, \bibinfo {author} {\bibfnamefont {D.}~\bibnamefont {Vaknin}}, \bibinfo
  {author} {\bibfnamefont {B.~C.}\ \bibnamefont {Sales}},\ and\ \bibinfo
  {author} {\bibfnamefont {R.~J.}\ \bibnamefont {McQueeney}},\ }\bibfield
  {title} {\bibinfo {title} {Crystal growth and magnetic structure of
  {M}n{B}i$_2${T}e$_4$},\ }\href@noop {} {\bibfield  {journal} {\bibinfo
  {journal} {Phys. Rev. Mater.}\ }\textbf {\bibinfo {volume} {3}},\ \bibinfo
  {pages} {064202} (\bibinfo {year} {2019})}\BibitemShut {NoStop}%
\bibitem [{\citenamefont {Gandhi}\ and\ \citenamefont
  {Ashby}(1983)}]{gandhi1983fracture}%
  \BibitemOpen
  \bibfield  {author} {\bibinfo {author} {\bibfnamefont {C.}~\bibnamefont
  {Gandhi}}\ and\ \bibinfo {author} {\bibfnamefont {M.~F.}\ \bibnamefont
  {Ashby}},\ }\bibfield  {title} {\bibinfo {title} {Fracture-mechanism maps for
  materials which cleave: {FCC}, {BCC} and {HCP} metals and ceramics},\
  }\href@noop {} {\bibfield  {journal} {\bibinfo  {journal} {Perspectives in
  Creep Fracture}\ ,\ \bibinfo {pages} {33}} (\bibinfo {year}
  {1983})}\BibitemShut {NoStop}%
\bibitem [{\citenamefont {Hernandez}\ \emph {et~al.}(2008)\citenamefont
  {Hernandez}, \citenamefont {Nicolosi}, \citenamefont {Lotya}, \citenamefont
  {Blighe}, \citenamefont {Sun}, \citenamefont {De}, \citenamefont {McGovern},
  \citenamefont {Holland}, \citenamefont {Byrne}, \citenamefont {Gun'Ko} \emph
  {et~al.}}]{hernandez2008high}%
  \BibitemOpen
  \bibfield  {author} {\bibinfo {author} {\bibfnamefont {Y.}~\bibnamefont
  {Hernandez}}, \bibinfo {author} {\bibfnamefont {V.}~\bibnamefont {Nicolosi}},
  \bibinfo {author} {\bibfnamefont {M.}~\bibnamefont {Lotya}}, \bibinfo
  {author} {\bibfnamefont {F.~M.}\ \bibnamefont {Blighe}}, \bibinfo {author}
  {\bibfnamefont {Z.}~\bibnamefont {Sun}}, \bibinfo {author} {\bibfnamefont
  {S.}~\bibnamefont {De}}, \bibinfo {author} {\bibfnamefont {I.~T.}\
  \bibnamefont {McGovern}}, \bibinfo {author} {\bibfnamefont {B.}~\bibnamefont
  {Holland}}, \bibinfo {author} {\bibfnamefont {M.}~\bibnamefont {Byrne}},
  \bibinfo {author} {\bibfnamefont {Y.~K.}\ \bibnamefont {Gun'Ko}}, \emph
  {et~al.},\ }\bibfield  {title} {\bibinfo {title} {High-yield production of
  graphene by liquid-phase exfoliation of graphite},\ }\href@noop {} {\bibfield
   {journal} {\bibinfo  {journal} {Nat. Nanotechnol.}\ }\textbf {\bibinfo
  {volume} {3}},\ \bibinfo {pages} {563} (\bibinfo {year} {2008})}\BibitemShut
  {NoStop}%
\bibitem [{\citenamefont {Ostendorf}\ \emph {et~al.}(2008)\citenamefont
  {Ostendorf}, \citenamefont {Schmitz}, \citenamefont {Hirth}, \citenamefont
  {K{\"u}hnle}, \citenamefont {Kolodziej},\ and\ \citenamefont
  {Reichling}}]{ostendorf2008flat}%
  \BibitemOpen
  \bibfield  {author} {\bibinfo {author} {\bibfnamefont {F.}~\bibnamefont
  {Ostendorf}}, \bibinfo {author} {\bibfnamefont {C.}~\bibnamefont {Schmitz}},
  \bibinfo {author} {\bibfnamefont {S.}~\bibnamefont {Hirth}}, \bibinfo
  {author} {\bibfnamefont {A.}~\bibnamefont {K{\"u}hnle}}, \bibinfo {author}
  {\bibfnamefont {J.~J.}\ \bibnamefont {Kolodziej}},\ and\ \bibinfo {author}
  {\bibfnamefont {M.}~\bibnamefont {Reichling}},\ }\bibfield  {title} {\bibinfo
  {title} {How flat is an air-cleaved mica surface?},\ }\href@noop {}
  {\bibfield  {journal} {\bibinfo  {journal} {Nanotechnology}\ }\textbf
  {\bibinfo {volume} {19}},\ \bibinfo {pages} {305705} (\bibinfo {year}
  {2008})}\BibitemShut {NoStop}%
\bibitem [{\citenamefont {Franceschi}\ \emph {et~al.}(2023)\citenamefont
  {Franceschi}, \citenamefont {Koc{\'a}n}, \citenamefont {Conti}, \citenamefont
  {Brandstetter}, \citenamefont {Balajka}, \citenamefont {Sokolovi{\'c}},
  \citenamefont {Valtiner}, \citenamefont {Mittendorfer}, \citenamefont
  {Schmid}, \citenamefont {Setv{\'\i}n} \emph
  {et~al.}}]{franceschi2023resolving}%
  \BibitemOpen
  \bibfield  {author} {\bibinfo {author} {\bibfnamefont {G.}~\bibnamefont
  {Franceschi}}, \bibinfo {author} {\bibfnamefont {P.}~\bibnamefont
  {Koc{\'a}n}}, \bibinfo {author} {\bibfnamefont {A.}~\bibnamefont {Conti}},
  \bibinfo {author} {\bibfnamefont {S.}~\bibnamefont {Brandstetter}}, \bibinfo
  {author} {\bibfnamefont {J.}~\bibnamefont {Balajka}}, \bibinfo {author}
  {\bibfnamefont {I.}~\bibnamefont {Sokolovi{\'c}}}, \bibinfo {author}
  {\bibfnamefont {M.}~\bibnamefont {Valtiner}}, \bibinfo {author}
  {\bibfnamefont {F.}~\bibnamefont {Mittendorfer}}, \bibinfo {author}
  {\bibfnamefont {M.}~\bibnamefont {Schmid}}, \bibinfo {author} {\bibfnamefont
  {M.}~\bibnamefont {Setv{\'\i}n}}, \emph {et~al.},\ }\bibfield  {title}
  {\bibinfo {title} {Resolving the intrinsic short-range ordering of {K}$^+$
  ions on cleaved muscovite mica},\ }\href@noop {} {\bibfield  {journal}
  {\bibinfo  {journal} {Nat. Commun.}\ }\textbf {\bibinfo {volume} {14}},\
  \bibinfo {pages} {208} (\bibinfo {year} {2023})}\BibitemShut {NoStop}%
\bibitem [{\citenamefont {Halwidl}\ \emph {et~al.}(2016)\citenamefont
  {Halwidl}, \citenamefont {St{\"o}ger}, \citenamefont {Mayr-Schm{\"o}lzer},
  \citenamefont {Pavelec}, \citenamefont {Fobes}, \citenamefont {Peng},
  \citenamefont {Mao}, \citenamefont {Schmid}, \citenamefont {Mittendorfer},
  \citenamefont {Redinger},\ and\ \citenamefont
  {Diebold}}]{halwidl2016adsorption}%
  \BibitemOpen
  \bibfield  {author} {\bibinfo {author} {\bibfnamefont {D.}~\bibnamefont
  {Halwidl}}, \bibinfo {author} {\bibfnamefont {B.}~\bibnamefont {St{\"o}ger}},
  \bibinfo {author} {\bibfnamefont {W.}~\bibnamefont {Mayr-Schm{\"o}lzer}},
  \bibinfo {author} {\bibfnamefont {J.}~\bibnamefont {Pavelec}}, \bibinfo
  {author} {\bibfnamefont {D.}~\bibnamefont {Fobes}}, \bibinfo {author}
  {\bibfnamefont {J.}~\bibnamefont {Peng}}, \bibinfo {author} {\bibfnamefont
  {G.~S.}\ \bibnamefont {Mao}, \bibfnamefont {Z.and~Parkinson}}, \bibinfo
  {author} {\bibfnamefont {M.}~\bibnamefont {Schmid}}, \bibinfo {author}
  {\bibfnamefont {F.}~\bibnamefont {Mittendorfer}}, \bibinfo {author}
  {\bibfnamefont {J.}~\bibnamefont {Redinger}},\ and\ \bibinfo {author}
  {\bibfnamefont {U.}~\bibnamefont {Diebold}},\ }\bibfield  {title} {\bibinfo
  {title} {Adsorption of water at the {SrO} surface of ruthenates},\
  }\href@noop {} {\bibfield  {journal} {\bibinfo  {journal} {Nat. Mater.}\
  }\textbf {\bibinfo {volume} {15}},\ \bibinfo {pages} {450} (\bibinfo {year}
  {2016})}\BibitemShut {NoStop}%
\bibitem [{\citenamefont {Ino}\ \emph {et~al.}(1964)\citenamefont {Ino},
  \citenamefont {Watanabe},\ and\ \citenamefont {Ogawa}}]{ino1964epitaxial}%
  \BibitemOpen
  \bibfield  {author} {\bibinfo {author} {\bibfnamefont {S.}~\bibnamefont
  {Ino}}, \bibinfo {author} {\bibfnamefont {D.}~\bibnamefont {Watanabe}},\ and\
  \bibinfo {author} {\bibfnamefont {S.}~\bibnamefont {Ogawa}},\ }\bibfield
  {title} {\bibinfo {title} {Epitaxial growth of metals on rocksalt faces
  cleaved in vacuum. {I}},\ }\href@noop {} {\bibfield  {journal} {\bibinfo
  {journal} {JPSJ}\ }\textbf {\bibinfo {volume} {19}},\ \bibinfo {pages} {881}
  (\bibinfo {year} {1964})}\BibitemShut {NoStop}%
\bibitem [{\citenamefont {Takagi}\ and\ \citenamefont
  {Hwang}(2010)}]{takagi2010emergent}%
  \BibitemOpen
  \bibfield  {author} {\bibinfo {author} {\bibfnamefont {H.}~\bibnamefont
  {Takagi}}\ and\ \bibinfo {author} {\bibfnamefont {H.~Y.}\ \bibnamefont
  {Hwang}},\ }\bibfield  {title} {\bibinfo {title} {An emergent change of phase
  for electronics},\ }\href@noop {} {\bibfield  {journal} {\bibinfo  {journal}
  {Science}\ }\textbf {\bibinfo {volume} {327}},\ \bibinfo {pages} {1601}
  (\bibinfo {year} {2010})}\BibitemShut {NoStop}%
\bibitem [{\citenamefont {No{\"e}l}\ \emph {et~al.}(2020)\citenamefont
  {No{\"e}l}, \citenamefont {Trier}, \citenamefont {Vicente~Arche},
  \citenamefont {Br{\'e}hin}, \citenamefont {Vaz}, \citenamefont {Garcia},
  \citenamefont {Fusil}, \citenamefont {Barth{\'e}l{\'e}my}, \citenamefont
  {Vila}, \citenamefont {Bibes} \emph {et~al.}}]{noel2020non}%
  \BibitemOpen
  \bibfield  {author} {\bibinfo {author} {\bibfnamefont {P.}~\bibnamefont
  {No{\"e}l}}, \bibinfo {author} {\bibfnamefont {F.}~\bibnamefont {Trier}},
  \bibinfo {author} {\bibfnamefont {L.~M.}\ \bibnamefont {Vicente~Arche}},
  \bibinfo {author} {\bibfnamefont {J.}~\bibnamefont {Br{\'e}hin}}, \bibinfo
  {author} {\bibfnamefont {D.~C.}\ \bibnamefont {Vaz}}, \bibinfo {author}
  {\bibfnamefont {V.}~\bibnamefont {Garcia}}, \bibinfo {author} {\bibfnamefont
  {S.}~\bibnamefont {Fusil}}, \bibinfo {author} {\bibfnamefont
  {A.}~\bibnamefont {Barth{\'e}l{\'e}my}}, \bibinfo {author} {\bibfnamefont
  {L.}~\bibnamefont {Vila}}, \bibinfo {author} {\bibfnamefont {M.}~\bibnamefont
  {Bibes}}, \emph {et~al.},\ }\bibfield  {title} {\bibinfo {title}
  {Non-volatile electric control of spin--charge conversion in a {SrTiO}$_3$
  rashba system},\ }\href@noop {} {\bibfield  {journal} {\bibinfo  {journal}
  {Nature}\ }\textbf {\bibinfo {volume} {580}},\ \bibinfo {pages} {483}
  (\bibinfo {year} {2020})}\BibitemShut {NoStop}%
\bibitem [{\citenamefont {Chen}\ \emph {et~al.}(2015)\citenamefont {Chen},
  \citenamefont {Trier}, \citenamefont {Wijnands}, \citenamefont {Green},
  \citenamefont {Gauquelin}, \citenamefont {Egoavil}, \citenamefont
  {Christensen}, \citenamefont {Koster}, \citenamefont {Huijben}, \citenamefont
  {Bovet} \emph {et~al.}}]{chen2015extreme}%
  \BibitemOpen
  \bibfield  {author} {\bibinfo {author} {\bibfnamefont {Y.~Z.}\ \bibnamefont
  {Chen}}, \bibinfo {author} {\bibfnamefont {F.}~\bibnamefont {Trier}},
  \bibinfo {author} {\bibfnamefont {T.}~\bibnamefont {Wijnands}}, \bibinfo
  {author} {\bibfnamefont {R.~J.}\ \bibnamefont {Green}}, \bibinfo {author}
  {\bibfnamefont {N.}~\bibnamefont {Gauquelin}}, \bibinfo {author}
  {\bibfnamefont {R.}~\bibnamefont {Egoavil}}, \bibinfo {author} {\bibfnamefont
  {D.~V.}\ \bibnamefont {Christensen}}, \bibinfo {author} {\bibfnamefont
  {G.}~\bibnamefont {Koster}}, \bibinfo {author} {\bibfnamefont
  {M.}~\bibnamefont {Huijben}}, \bibinfo {author} {\bibfnamefont
  {N.}~\bibnamefont {Bovet}}, \emph {et~al.},\ }\bibfield  {title} {\bibinfo
  {title} {Extreme mobility enhancement of two-dimensional electron gases at
  oxide interfaces by charge-transfer-induced modulation doping},\ }\href@noop
  {} {\bibfield  {journal} {\bibinfo  {journal} {Nat. Mater.}\ }\textbf
  {\bibinfo {volume} {14}},\ \bibinfo {pages} {801} (\bibinfo {year}
  {2015})}\BibitemShut {NoStop}%
\bibitem [{\citenamefont {Varignon}\ \emph {et~al.}(2018)\citenamefont
  {Varignon}, \citenamefont {Vila}, \citenamefont {Barth{\'e}l{\'e}my},\ and\
  \citenamefont {Bibes}}]{varignon2018new}%
  \BibitemOpen
  \bibfield  {author} {\bibinfo {author} {\bibfnamefont {J.}~\bibnamefont
  {Varignon}}, \bibinfo {author} {\bibfnamefont {L.}~\bibnamefont {Vila}},
  \bibinfo {author} {\bibfnamefont {A.}~\bibnamefont {Barth{\'e}l{\'e}my}},\
  and\ \bibinfo {author} {\bibfnamefont {M.}~\bibnamefont {Bibes}},\ }\bibfield
   {title} {\bibinfo {title} {A new spin for oxide interfaces},\ }\href@noop {}
  {\bibfield  {journal} {\bibinfo  {journal} {Nat. Phys.}\ }\textbf {\bibinfo
  {volume} {14}},\ \bibinfo {pages} {322} (\bibinfo {year} {2018})}\BibitemShut
  {NoStop}%
\bibitem [{\citenamefont {Wang}\ \emph {et~al.}(2021)\citenamefont {Wang},
  \citenamefont {Han}, \citenamefont {Shao}, \citenamefont {Qiu}, \citenamefont
  {Wang},\ and\ \citenamefont {Liu}}]{wang2021perovskite}%
  \BibitemOpen
  \bibfield  {author} {\bibinfo {author} {\bibfnamefont {K.}~\bibnamefont
  {Wang}}, \bibinfo {author} {\bibfnamefont {C.}~\bibnamefont {Han}}, \bibinfo
  {author} {\bibfnamefont {Z.}~\bibnamefont {Shao}}, \bibinfo {author}
  {\bibfnamefont {J.}~\bibnamefont {Qiu}}, \bibinfo {author} {\bibfnamefont
  {S.}~\bibnamefont {Wang}},\ and\ \bibinfo {author} {\bibfnamefont
  {S.}~\bibnamefont {Liu}},\ }\bibfield  {title} {\bibinfo {title} {Perovskite
  oxide catalysts for advanced oxidation reactions},\ }\href@noop {} {\bibfield
   {journal} {\bibinfo  {journal} {Adv. Func. Mater.}\ }\textbf {\bibinfo
  {volume} {31}},\ \bibinfo {pages} {2102089} (\bibinfo {year}
  {2021})}\BibitemShut {NoStop}%
\bibitem [{\citenamefont {Jun}\ \emph {et~al.}(2016)\citenamefont {Jun},
  \citenamefont {Kim}, \citenamefont {Shin},\ and\ \citenamefont
  {Kim}}]{jun2016perovskite}%
  \BibitemOpen
  \bibfield  {author} {\bibinfo {author} {\bibfnamefont {A.}~\bibnamefont
  {Jun}}, \bibinfo {author} {\bibfnamefont {J.}~\bibnamefont {Kim}}, \bibinfo
  {author} {\bibfnamefont {J.}~\bibnamefont {Shin}},\ and\ \bibinfo {author}
  {\bibfnamefont {G.}~\bibnamefont {Kim}},\ }\bibfield  {title} {\bibinfo
  {title} {Perovskite as a cathode material: a review of its role in
  solid-oxide fuel cell technology},\ }\href@noop {} {\bibfield  {journal}
  {\bibinfo  {journal} {ChemElectroChem}\ }\textbf {\bibinfo {volume} {3}},\
  \bibinfo {pages} {511} (\bibinfo {year} {2016})}\BibitemShut {NoStop}%
\bibitem [{\citenamefont {Shu}\ \emph {et~al.}(2019)\citenamefont {Shu},
  \citenamefont {Sunarso}, \citenamefont {Hashim}, \citenamefont {Mao},
  \citenamefont {Zhou},\ and\ \citenamefont {Liang}}]{shu2019advanced}%
  \BibitemOpen
  \bibfield  {author} {\bibinfo {author} {\bibfnamefont {L.}~\bibnamefont
  {Shu}}, \bibinfo {author} {\bibfnamefont {J.}~\bibnamefont {Sunarso}},
  \bibinfo {author} {\bibfnamefont {S.~S.}\ \bibnamefont {Hashim}}, \bibinfo
  {author} {\bibfnamefont {J.}~\bibnamefont {Mao}}, \bibinfo {author}
  {\bibfnamefont {W.}~\bibnamefont {Zhou}},\ and\ \bibinfo {author}
  {\bibfnamefont {F.}~\bibnamefont {Liang}},\ }\bibfield  {title} {\bibinfo
  {title} {Advanced perovskite anodes for solid oxide fuel cells: A review},\
  }\href@noop {} {\bibfield  {journal} {\bibinfo  {journal} {Int. J. Hydrog.
  Energy}\ }\textbf {\bibinfo {volume} {44}},\ \bibinfo {pages} {31275}
  (\bibinfo {year} {2019})}\BibitemShut {NoStop}%
\bibitem [{\citenamefont {Sokolovi{\'c}}\ \emph {et~al.}(2019)\citenamefont
  {Sokolovi{\'c}}, \citenamefont {Schmid}, \citenamefont {Diebold},\ and\
  \citenamefont {Setvín}}]{sokolovic2019incipient}%
  \BibitemOpen
  \bibfield  {author} {\bibinfo {author} {\bibfnamefont {I.}~\bibnamefont
  {Sokolovi{\'c}}}, \bibinfo {author} {\bibfnamefont {M.}~\bibnamefont
  {Schmid}}, \bibinfo {author} {\bibfnamefont {U.}~\bibnamefont {Diebold}},\
  and\ \bibinfo {author} {\bibfnamefont {M.}~\bibnamefont {Setvín}},\
  }\bibfield  {title} {\bibinfo {title} {Incipient ferroelectricity: A route
  towards bulk-terminated {SrTiO}$_3$},\ }\href@noop {} {\bibfield  {journal}
  {\bibinfo  {journal} {Phys. Rev. Mater.}\ }\textbf {\bibinfo {volume} {3}},\
  \bibinfo {pages} {034407} (\bibinfo {year} {2019})}\BibitemShut {NoStop}%
\bibitem [{\citenamefont {Sokolovi{\'c}}\ \emph {et~al.}(2021)\citenamefont
  {Sokolovi{\'c}}, \citenamefont {Franceschi}, \citenamefont {Wang},
  \citenamefont {Xu}, \citenamefont {Pavelec}, \citenamefont {Riva},
  \citenamefont {Schmid}, \citenamefont {Diebold},\ and\ \citenamefont
  {Setv{\'\i}n}}]{sokolovic2021quest}%
  \BibitemOpen
  \bibfield  {author} {\bibinfo {author} {\bibfnamefont {I.}~\bibnamefont
  {Sokolovi{\'c}}}, \bibinfo {author} {\bibfnamefont {G.}~\bibnamefont
  {Franceschi}}, \bibinfo {author} {\bibfnamefont {Z.}~\bibnamefont {Wang}},
  \bibinfo {author} {\bibfnamefont {J.}~\bibnamefont {Xu}}, \bibinfo {author}
  {\bibfnamefont {J.}~\bibnamefont {Pavelec}}, \bibinfo {author} {\bibfnamefont
  {M.}~\bibnamefont {Riva}}, \bibinfo {author} {\bibfnamefont {M.}~\bibnamefont
  {Schmid}}, \bibinfo {author} {\bibfnamefont {U.}~\bibnamefont {Diebold}},\
  and\ \bibinfo {author} {\bibfnamefont {M.}~\bibnamefont {Setv{\'\i}n}},\
  }\bibfield  {title} {\bibinfo {title} {Quest for a pristine unreconstructed
  {SrTiO}$_3$(001) surface: {A}n atomically resolved study via noncontact
  atomic force microscopy},\ }\href@noop {} {\bibfield  {journal} {\bibinfo
  {journal} {Phys. Rev. B}\ }\textbf {\bibinfo {volume} {103}},\ \bibinfo
  {pages} {L241406} (\bibinfo {year} {2021})}\BibitemShut {NoStop}%
\bibitem [{\citenamefont {M{\"u}ller}\ \emph {et~al.}(1968)\citenamefont
  {M{\"u}ller}, \citenamefont {Berlinger},\ and\ \citenamefont
  {Waldner}}]{muller1968characteristic}%
  \BibitemOpen
  \bibfield  {author} {\bibinfo {author} {\bibfnamefont {K.~A.}\ \bibnamefont
  {M{\"u}ller}}, \bibinfo {author} {\bibfnamefont {W.}~\bibnamefont
  {Berlinger}},\ and\ \bibinfo {author} {\bibfnamefont {F.}~\bibnamefont
  {Waldner}},\ }\bibfield  {title} {\bibinfo {title} {Characteristic structural
  phase transition in perovskite-type compounds},\ }\href@noop {} {\bibfield
  {journal} {\bibinfo  {journal} {Phys. Rev. Lett.}\ }\textbf {\bibinfo
  {volume} {21}},\ \bibinfo {pages} {814} (\bibinfo {year} {1968})}\BibitemShut
  {NoStop}%
\bibitem [{\citenamefont {M{\"u}ller}\ and\ \citenamefont
  {Burkard}(1979)}]{muller1979srtio3}%
  \BibitemOpen
  \bibfield  {author} {\bibinfo {author} {\bibfnamefont {K.~A.}\ \bibnamefont
  {M{\"u}ller}}\ and\ \bibinfo {author} {\bibfnamefont {H.}~\bibnamefont
  {Burkard}},\ }\bibfield  {title} {\bibinfo {title} {{S}r{T}i{O}$_3$: {A}n
  intrinsic quantum paraelectric below 4 {K}},\ }\href@noop {} {\bibfield
  {journal} {\bibinfo  {journal} {Phys. Rev. B}\ }\textbf {\bibinfo {volume}
  {19}},\ \bibinfo {pages} {3593} (\bibinfo {year} {1979})}\BibitemShut
  {NoStop}%
\bibitem [{\citenamefont {Burke}\ and\ \citenamefont
  {Pressley}(1971)}]{burke1971stress}%
  \BibitemOpen
  \bibfield  {author} {\bibinfo {author} {\bibfnamefont {W.~J.}\ \bibnamefont
  {Burke}}\ and\ \bibinfo {author} {\bibfnamefont {R.~J.}\ \bibnamefont
  {Pressley}},\ }\bibfield  {title} {\bibinfo {title} {Stress induced
  ferroelectricity in {S}r{T}i{O}$_3$},\ }\href@noop {} {\bibfield  {journal}
  {\bibinfo  {journal} {Solid State Commun.}\ }\textbf {\bibinfo {volume}
  {9}},\ \bibinfo {pages} {191} (\bibinfo {year} {1971})}\BibitemShut {NoStop}%
\bibitem [{\citenamefont {Haeni}\ \emph {et~al.}(2004)\citenamefont {Haeni},
  \citenamefont {Irvin}, \citenamefont {Chang}, \citenamefont {Uecker},
  \citenamefont {Reiche}, \citenamefont {Li}, \citenamefont {Choudhury},
  \citenamefont {Tian}, \citenamefont {Hawley}, \citenamefont {Craigo},
  \citenamefont {Tagantsev}, \citenamefont {Pan}, \citenamefont {Streiffer},
  \citenamefont {Chen}, \citenamefont {Kirchoefer}, \citenamefont {Levy},\ and\
  \citenamefont {Schlom}}]{haeni2004room}%
  \BibitemOpen
  \bibfield  {author} {\bibinfo {author} {\bibfnamefont {J.~H.}\ \bibnamefont
  {Haeni}}, \bibinfo {author} {\bibfnamefont {P.}~\bibnamefont {Irvin}},
  \bibinfo {author} {\bibfnamefont {W.}~\bibnamefont {Chang}}, \bibinfo
  {author} {\bibfnamefont {R.}~\bibnamefont {Uecker}}, \bibinfo {author}
  {\bibfnamefont {P.}~\bibnamefont {Reiche}}, \bibinfo {author} {\bibfnamefont
  {Y.~L.}\ \bibnamefont {Li}}, \bibinfo {author} {\bibfnamefont
  {S.}~\bibnamefont {Choudhury}}, \bibinfo {author} {\bibfnamefont
  {W.}~\bibnamefont {Tian}}, \bibinfo {author} {\bibfnamefont {M.~E.}\
  \bibnamefont {Hawley}}, \bibinfo {author} {\bibfnamefont {B.}~\bibnamefont
  {Craigo}}, \bibinfo {author} {\bibfnamefont {A.~K.}\ \bibnamefont
  {Tagantsev}}, \bibinfo {author} {\bibfnamefont {X.~Q.}\ \bibnamefont {Pan}},
  \bibinfo {author} {\bibfnamefont {S.~K.}\ \bibnamefont {Streiffer}}, \bibinfo
  {author} {\bibfnamefont {L.~Q.}\ \bibnamefont {Chen}}, \bibinfo {author}
  {\bibfnamefont {S.~W.}\ \bibnamefont {Kirchoefer}}, \bibinfo {author}
  {\bibfnamefont {J.}~\bibnamefont {Levy}},\ and\ \bibinfo {author}
  {\bibfnamefont {D.~G.}\ \bibnamefont {Schlom}},\ }\bibfield  {title}
  {\bibinfo {title} {Room-temperature ferroelectricity in strained
  {S}r{TiO}$_3$},\ }\href@noop {} {\bibfield  {journal} {\bibinfo  {journal}
  {Nature}\ }\textbf {\bibinfo {volume} {430}},\ \bibinfo {pages} {758}
  (\bibinfo {year} {2004})}\BibitemShut {NoStop}%
\bibitem [{\citenamefont {Griffith}(1921)}]{griffith1921phenomena}%
  \BibitemOpen
  \bibfield  {author} {\bibinfo {author} {\bibfnamefont {A.~A.}\ \bibnamefont
  {Griffith}},\ }\bibfield  {title} {\bibinfo {title} {The phenomena of rupture
  and flow in solids},\ }\href@noop {} {\bibfield  {journal} {\bibinfo
  {journal} {Philos. Trans. R. Soc. A}\ }\textbf {\bibinfo {volume} {221}},\
  \bibinfo {pages} {163} (\bibinfo {year} {1921})}\BibitemShut {NoStop}%
\bibitem [{\citenamefont {Meza}\ \emph {et~al.}(2024)\citenamefont {Meza},
  \citenamefont {Baddoo},\ and\ \citenamefont {Gardner}}]{Meza2024}%
  \BibitemOpen
  \bibfield  {author} {\bibinfo {author} {\bibfnamefont {F.~J.}\ \bibnamefont
  {Meza}}, \bibinfo {author} {\bibfnamefont {N.}~\bibnamefont {Baddoo}},\ and\
  \bibinfo {author} {\bibfnamefont {L.}~\bibnamefont {Gardner}},\ }\bibfield
  {title} {\bibinfo {title} {Derivation of stainless steel material factors for
  {E}uropean and {U.S.} design standards},\ }\href@noop {} {\bibfield
  {journal} {\bibinfo  {journal} {J. Construct. Steel Res.}\ }\textbf {\bibinfo
  {volume} {213}},\ \bibinfo {pages} {108383} (\bibinfo {year}
  {2024})}\BibitemShut {NoStop}%
\bibitem [{\citenamefont {Setv{\'\i}n}\ \emph {et~al.}(2012)\citenamefont
  {Setv{\'\i}n}, \citenamefont {Javorsk{\`y}}, \citenamefont
  {Tur{\v{c}}inkov{\'a}}, \citenamefont {Matol{\'\i}nov{\'a}}, \citenamefont
  {Sobot{\'\i}k}, \citenamefont {Koc{\'a}n},\ and\ \citenamefont
  {O{\v{s}}t’{\'a}dal}}]{setvin2012ultrasharp}%
  \BibitemOpen
  \bibfield  {author} {\bibinfo {author} {\bibfnamefont {M.}~\bibnamefont
  {Setv{\'\i}n}}, \bibinfo {author} {\bibfnamefont {J.}~\bibnamefont
  {Javorsk{\`y}}}, \bibinfo {author} {\bibfnamefont {D.}~\bibnamefont
  {Tur{\v{c}}inkov{\'a}}}, \bibinfo {author} {\bibfnamefont {I.}~\bibnamefont
  {Matol{\'\i}nov{\'a}}}, \bibinfo {author} {\bibfnamefont {P.}~\bibnamefont
  {Sobot{\'\i}k}}, \bibinfo {author} {\bibfnamefont {P.}~\bibnamefont
  {Koc{\'a}n}},\ and\ \bibinfo {author} {\bibfnamefont {I.}~\bibnamefont
  {O{\v{s}}t’{\'a}dal}},\ }\bibfield  {title} {\bibinfo {title} {Ultrasharp
  tungsten tips—characterization and nondestructive cleaning},\ }\href@noop
  {} {\bibfield  {journal} {\bibinfo  {journal} {Ultramicroscopy}\ }\textbf
  {\bibinfo {volume} {113}},\ \bibinfo {pages} {152} (\bibinfo {year}
  {2012})}\BibitemShut {NoStop}%
\bibitem [{\citenamefont {Giessibl}(2019)}]{Giessibl2019Review}%
  \BibitemOpen
  \bibfield  {author} {\bibinfo {author} {\bibfnamefont {F.~J.}\ \bibnamefont
  {Giessibl}},\ }\bibfield  {title} {\bibinfo {title} {The q{P}lus sensor, a
  powerful core for the atomic force microscope},\ }\href@noop {} {\bibfield
  {journal} {\bibinfo  {journal} {Rev. Sci. Instr.}\ }\textbf {\bibinfo
  {volume} {90}},\ \bibinfo {pages} {011101} (\bibinfo {year}
  {2019})}\BibitemShut {NoStop}%
\bibitem [{\citenamefont {Huber}\ and\ \citenamefont
  {Giessibl}(2017)}]{huber2017low}%
  \BibitemOpen
  \bibfield  {author} {\bibinfo {author} {\bibfnamefont {F.}~\bibnamefont
  {Huber}}\ and\ \bibinfo {author} {\bibfnamefont {F.~J.}\ \bibnamefont
  {Giessibl}},\ }\bibfield  {title} {\bibinfo {title} {Low noise current
  preamplifier for {qPlus} sensor deflection signal detection in atomic force
  microscopy at room and low temperatures},\ }\href@noop {} {\bibfield
  {journal} {\bibinfo  {journal} {Rev. of Sci. Instr.}\ }\textbf {\bibinfo
  {volume} {88}},\ \bibinfo {pages} {073702} (\bibinfo {year}
  {2017})}\BibitemShut {NoStop}%
\bibitem [{\citenamefont {Schmid}\ \emph {et~al.}(2019)\citenamefont {Schmid},
  \citenamefont {Setvín},\ and\ \citenamefont {Diebold}}]{schmid2019device}%
  \BibitemOpen
  \bibfield  {author} {\bibinfo {author} {\bibfnamefont {M.}~\bibnamefont
  {Schmid}}, \bibinfo {author} {\bibfnamefont {M.}~\bibnamefont {Setvín}},\
  and\ \bibinfo {author} {\bibfnamefont {U.}~\bibnamefont {Diebold}},\
  }\href@noop {} {\bibinfo {title} {Device for suspending a load in a
  vibration-insulated manner}} (\bibinfo {year} {2019}),\ \bibinfo {note} {{US}
  Patent App. 16/327,528}\BibitemShut {NoStop}%
\bibitem [{\citenamefont {Sokolovi{\'c}}\ \emph {et~al.}(2020)\citenamefont
  {Sokolovi{\'c}}, \citenamefont {Reticcioli}, \citenamefont
  {{\v{C}}alkovsk{\`y}}, \citenamefont {Wagner}, \citenamefont {Schmid},
  \citenamefont {Franchini}, \citenamefont {Diebold},\ and\ \citenamefont
  {Setv{\'\i}n}}]{sokolovic2020resolving}%
  \BibitemOpen
  \bibfield  {author} {\bibinfo {author} {\bibfnamefont {I.}~\bibnamefont
  {Sokolovi{\'c}}}, \bibinfo {author} {\bibfnamefont {M.}~\bibnamefont
  {Reticcioli}}, \bibinfo {author} {\bibfnamefont {M.}~\bibnamefont
  {{\v{C}}alkovsk{\`y}}}, \bibinfo {author} {\bibfnamefont {M.}~\bibnamefont
  {Wagner}}, \bibinfo {author} {\bibfnamefont {M.}~\bibnamefont {Schmid}},
  \bibinfo {author} {\bibfnamefont {C.}~\bibnamefont {Franchini}}, \bibinfo
  {author} {\bibfnamefont {U.}~\bibnamefont {Diebold}},\ and\ \bibinfo {author}
  {\bibfnamefont {M.}~\bibnamefont {Setv{\'\i}n}},\ }\bibfield  {title}
  {\bibinfo {title} {Resolving the adsorption of molecular {O}$_2$ on the
  rutile {T}i{O}$_2$(110) surface by noncontact atomic force microscopy},\
  }\href@noop {} {\bibfield  {journal} {\bibinfo  {journal} {PNAS}\ }\textbf
  {\bibinfo {volume} {117}},\ \bibinfo {pages} {14827} (\bibinfo {year}
  {2020})}\BibitemShut {NoStop}%
\bibitem [{\citenamefont {Chien}\ \emph {et~al.}(2009)\citenamefont {Chien},
  \citenamefont {Santos}, \citenamefont {Bode}, \citenamefont {Guisinger},\
  and\ \citenamefont {Freeland}}]{chien2009controllable}%
  \BibitemOpen
  \bibfield  {author} {\bibinfo {author} {\bibfnamefont {T.-Y.}\ \bibnamefont
  {Chien}}, \bibinfo {author} {\bibfnamefont {T.~S.}\ \bibnamefont {Santos}},
  \bibinfo {author} {\bibfnamefont {M.}~\bibnamefont {Bode}}, \bibinfo {author}
  {\bibfnamefont {N.~P.}\ \bibnamefont {Guisinger}},\ and\ \bibinfo {author}
  {\bibfnamefont {J.~W.}\ \bibnamefont {Freeland}},\ }\bibfield  {title}
  {\bibinfo {title} {Controllable local modification of fractured {N}b-doped
  {S}r{T}i{O}$_3$ surfaces},\ }\href@noop {} {\bibfield  {journal} {\bibinfo
  {journal} {Appl. Phys. Lett.}\ }\textbf {\bibinfo {volume} {95}},\ \bibinfo
  {pages} {163107} (\bibinfo {year} {2009})}\BibitemShut {NoStop}%
\bibitem [{\citenamefont {Chien}\ \emph {et~al.}(2010)\citenamefont {Chien},
  \citenamefont {Guisinger},\ and\ \citenamefont {Freeland}}]{chien2010survey}%
  \BibitemOpen
  \bibfield  {author} {\bibinfo {author} {\bibfnamefont {T.-Y.}\ \bibnamefont
  {Chien}}, \bibinfo {author} {\bibfnamefont {N.~P.}\ \bibnamefont
  {Guisinger}},\ and\ \bibinfo {author} {\bibfnamefont {J.~W.}\ \bibnamefont
  {Freeland}},\ }\bibfield  {title} {\bibinfo {title} {Survey of fractured
  {S}r{T}i{O}$_3$ surfaces: from the micrometer to nanometer scale},\
  }\href@noop {} {\bibfield  {journal} {\bibinfo  {journal} {J. Vac. Sci.
  Technol. B}\ }\textbf {\bibinfo {volume} {28}},\ \bibinfo {pages} {C5A11}
  (\bibinfo {year} {2010})}\BibitemShut {NoStop}%
\bibitem [{\citenamefont {Guisinger}\ \emph {et~al.}(2009)\citenamefont
  {Guisinger}, \citenamefont {Santos}, \citenamefont {Guest}, \citenamefont
  {Chien}, \citenamefont {Bhattacharya}, \citenamefont {Freeland},\ and\
  \citenamefont {Bode}}]{guisinger2009nanometer}%
  \BibitemOpen
  \bibfield  {author} {\bibinfo {author} {\bibfnamefont {N.~P.}\ \bibnamefont
  {Guisinger}}, \bibinfo {author} {\bibfnamefont {T.~S.}\ \bibnamefont
  {Santos}}, \bibinfo {author} {\bibfnamefont {J.~R.}\ \bibnamefont {Guest}},
  \bibinfo {author} {\bibfnamefont {T.-Y.}\ \bibnamefont {Chien}}, \bibinfo
  {author} {\bibfnamefont {A.}~\bibnamefont {Bhattacharya}}, \bibinfo {author}
  {\bibfnamefont {J.~W.}\ \bibnamefont {Freeland}},\ and\ \bibinfo {author}
  {\bibfnamefont {M.}~\bibnamefont {Bode}},\ }\bibfield  {title} {\bibinfo
  {title} {Nanometer-scale striped surface terminations on fractured
  {S}r{T}i{O}$_3$ surfaces},\ }\href@noop {} {\bibfield  {journal} {\bibinfo
  {journal} {ACS Nano}\ }\textbf {\bibinfo {volume} {3}},\ \bibinfo {pages}
  {4132} (\bibinfo {year} {2009})}\BibitemShut {NoStop}%
\bibitem [{\citenamefont {Sitaputra}\ \emph {et~al.}(2015)\citenamefont
  {Sitaputra}, \citenamefont {Skowronski},\ and\ \citenamefont
  {Feenstra}}]{sitaputra2015topographic}%
  \BibitemOpen
  \bibfield  {author} {\bibinfo {author} {\bibfnamefont {W.}~\bibnamefont
  {Sitaputra}}, \bibinfo {author} {\bibfnamefont {M.}~\bibnamefont
  {Skowronski}},\ and\ \bibinfo {author} {\bibfnamefont {R.~M.}\ \bibnamefont
  {Feenstra}},\ }\bibfield  {title} {\bibinfo {title} {Topographic and
  electronic structure of cleaved {S}r{T}i{O}$_3$ (001) surfaces},\ }\href@noop
  {} {\bibfield  {journal} {\bibinfo  {journal} {J. Vac. Sci. Technol. A}\
  }\textbf {\bibinfo {volume} {33}},\ \bibinfo {pages} {031402} (\bibinfo
  {year} {2015})}\BibitemShut {NoStop}%
\bibitem [{\citenamefont {Hunter}\ \emph {et~al.}(2024)\citenamefont {Hunter},
  \citenamefont {Putzke}, \citenamefont {Gaponenko}, \citenamefont {Tamai},
  \citenamefont {Baumberger},\ and\ \citenamefont
  {Moll}}]{hunter2024controlling}%
  \BibitemOpen
  \bibfield  {author} {\bibinfo {author} {\bibfnamefont {A.}~\bibnamefont
  {Hunter}}, \bibinfo {author} {\bibfnamefont {C.}~\bibnamefont {Putzke}},
  \bibinfo {author} {\bibfnamefont {I.}~\bibnamefont {Gaponenko}}, \bibinfo
  {author} {\bibfnamefont {A.}~\bibnamefont {Tamai}}, \bibinfo {author}
  {\bibfnamefont {F.}~\bibnamefont {Baumberger}},\ and\ \bibinfo {author}
  {\bibfnamefont {P.~J.~W.}\ \bibnamefont {Moll}},\ }\bibfield  {title}
  {\bibinfo {title} {Controlling crystal cleavage in focused ion beam shaped
  specimens for surface spectroscopy},\ }\href@noop {} {\bibfield  {journal}
  {\bibinfo  {journal} {Rev. of Sci. Instr.}\ }\textbf {\bibinfo {volume} {95}}
  (\bibinfo {year} {2024})}\BibitemShut {NoStop}%
\bibitem [{\citenamefont {Sakudo}\ and\ \citenamefont
  {Unoki}(1971)}]{sakudo1971dielectric}%
  \BibitemOpen
  \bibfield  {author} {\bibinfo {author} {\bibfnamefont {T.}~\bibnamefont
  {Sakudo}}\ and\ \bibinfo {author} {\bibfnamefont {H.}~\bibnamefont {Unoki}},\
  }\bibfield  {title} {\bibinfo {title} {Dielectric properties of
  {S}r{T}i{O}$_3$ at low temperatures},\ }\href@noop {} {\bibfield  {journal}
  {\bibinfo  {journal} {Phys. Rev. Lett.}\ }\textbf {\bibinfo {volume} {26}},\
  \bibinfo {pages} {851} (\bibinfo {year} {1971})}\BibitemShut {NoStop}%
\bibitem [{\citenamefont {Aschauer}\ and\ \citenamefont
  {Spaldin}(2014)}]{aschauer2014competition}%
  \BibitemOpen
  \bibfield  {author} {\bibinfo {author} {\bibfnamefont {U.}~\bibnamefont
  {Aschauer}}\ and\ \bibinfo {author} {\bibfnamefont {N.~A.}\ \bibnamefont
  {Spaldin}},\ }\bibfield  {title} {\bibinfo {title} {Competition and
  cooperation between antiferrodistortive and ferroelectric instabilities in
  the model perovskite {S}r{T}i{O}$_3$},\ }\href@noop {} {\bibfield  {journal}
  {\bibinfo  {journal} {Journal of Physics: Condensed Matter}\ }\textbf
  {\bibinfo {volume} {26}},\ \bibinfo {pages} {122203} (\bibinfo {year}
  {2014})}\BibitemShut {NoStop}%
\bibitem [{\citenamefont {Kawasaki}\ \emph {et~al.}(1994)\citenamefont
  {Kawasaki}, \citenamefont {Takahashi}, \citenamefont {Maeda}, \citenamefont
  {Tsuchiya}, \citenamefont {Shinohara}, \citenamefont {Ishiyama},
  \citenamefont {Yonezawa}, \citenamefont {Yoshimoto},\ and\ \citenamefont
  {Koinuma}}]{kawasaki1994atomic}%
  \BibitemOpen
  \bibfield  {author} {\bibinfo {author} {\bibfnamefont {M.}~\bibnamefont
  {Kawasaki}}, \bibinfo {author} {\bibfnamefont {K.}~\bibnamefont {Takahashi}},
  \bibinfo {author} {\bibfnamefont {T.}~\bibnamefont {Maeda}}, \bibinfo
  {author} {\bibfnamefont {R.}~\bibnamefont {Tsuchiya}}, \bibinfo {author}
  {\bibfnamefont {M.}~\bibnamefont {Shinohara}}, \bibinfo {author}
  {\bibfnamefont {O.}~\bibnamefont {Ishiyama}}, \bibinfo {author}
  {\bibfnamefont {T.}~\bibnamefont {Yonezawa}}, \bibinfo {author}
  {\bibfnamefont {M.}~\bibnamefont {Yoshimoto}},\ and\ \bibinfo {author}
  {\bibfnamefont {H.}~\bibnamefont {Koinuma}},\ }\bibfield  {title} {\bibinfo
  {title} {Atomic control of the {SrTiO}$_3$ crystal surface},\ }\href@noop {}
  {\bibfield  {journal} {\bibinfo  {journal} {Science}\ }\textbf {\bibinfo
  {volume} {266}},\ \bibinfo {pages} {1540} (\bibinfo {year}
  {1994})}\BibitemShut {NoStop}%
\bibitem [{\citenamefont {Connell}\ \emph {et~al.}(2012)\citenamefont
  {Connell}, \citenamefont {Isaac}, \citenamefont {Ekanayake}, \citenamefont
  {Strachan},\ and\ \citenamefont {Seo}}]{connell2012preparation}%
  \BibitemOpen
  \bibfield  {author} {\bibinfo {author} {\bibfnamefont {J.~G.}\ \bibnamefont
  {Connell}}, \bibinfo {author} {\bibfnamefont {B.~J.}\ \bibnamefont {Isaac}},
  \bibinfo {author} {\bibfnamefont {G.~B.}\ \bibnamefont {Ekanayake}}, \bibinfo
  {author} {\bibfnamefont {D.~R.}\ \bibnamefont {Strachan}},\ and\ \bibinfo
  {author} {\bibfnamefont {S.~S.~A.}\ \bibnamefont {Seo}},\ }\bibfield  {title}
  {\bibinfo {title} {Preparation of atomically flat {SrTiO}$_3$ surfaces using
  a deionized-water leaching and thermal annealing procedure},\ }\href@noop {}
  {\bibfield  {journal} {\bibinfo  {journal} {Appl. Phys. Lett.}\ }\textbf
  {\bibinfo {volume} {101}},\ \bibinfo {pages} {251607} (\bibinfo {year}
  {2012})}\BibitemShut {NoStop}%
\bibitem [{\citenamefont {Deak}\ \emph {et~al.}(2006)\citenamefont {Deak},
  \citenamefont {Silly}, \citenamefont {Newell},\ and\ \citenamefont
  {Castell}}]{deak2006ordering}%
  \BibitemOpen
  \bibfield  {author} {\bibinfo {author} {\bibfnamefont {D.~S.}\ \bibnamefont
  {Deak}}, \bibinfo {author} {\bibfnamefont {F.}~\bibnamefont {Silly}},
  \bibinfo {author} {\bibfnamefont {D.~T.}\ \bibnamefont {Newell}},\ and\
  \bibinfo {author} {\bibfnamefont {M.~R.}\ \bibnamefont {Castell}},\
  }\bibfield  {title} {\bibinfo {title} {Ordering of {TiO}$_2$-based
  nanostructures on {SrTiO}$_3$(001) surfaces},\ }\href@noop {} {\bibfield
  {journal} {\bibinfo  {journal} {J. Phys. Chem. B}\ }\textbf {\bibinfo
  {volume} {110}},\ \bibinfo {pages} {9246} (\bibinfo {year}
  {2006})}\BibitemShut {NoStop}%
\bibitem [{\citenamefont {Chambers}\ \emph {et~al.}(2012)\citenamefont
  {Chambers}, \citenamefont {Droubay}, \citenamefont {Capan},\ and\
  \citenamefont {Sun}}]{chambers2012unintentional}%
  \BibitemOpen
  \bibfield  {author} {\bibinfo {author} {\bibfnamefont {S.~A.}\ \bibnamefont
  {Chambers}}, \bibinfo {author} {\bibfnamefont {T.~C.}\ \bibnamefont
  {Droubay}}, \bibinfo {author} {\bibfnamefont {C.}~\bibnamefont {Capan}},\
  and\ \bibinfo {author} {\bibfnamefont {G.~Y.}\ \bibnamefont {Sun}},\
  }\bibfield  {title} {\bibinfo {title} {Unintentional {F} doping of
  {SrTiO}$_3$ (001) etched in {HF} acid-structure and electronic properties},\
  }\href@noop {} {\bibfield  {journal} {\bibinfo  {journal} {Surf. Sci.}\
  }\textbf {\bibinfo {volume} {606}},\ \bibinfo {pages} {554} (\bibinfo {year}
  {2012})}\BibitemShut {NoStop}%
\bibitem [{\citenamefont {Castell}(2002)}]{castell2002scanning}%
  \BibitemOpen
  \bibfield  {author} {\bibinfo {author} {\bibfnamefont {M.~R.}\ \bibnamefont
  {Castell}},\ }\bibfield  {title} {\bibinfo {title} {Scanning tunneling
  microscopy of reconstructions on the {SrTiO}$_3$(001) surface},\ }\href@noop
  {} {\bibfield  {journal} {\bibinfo  {journal} {Surf. Sci.}\ }\textbf
  {\bibinfo {volume} {505}},\ \bibinfo {pages} {1} (\bibinfo {year}
  {2002})}\BibitemShut {NoStop}%
\bibitem [{\citenamefont {Silly}\ \emph {et~al.}(2006)\citenamefont {Silly},
  \citenamefont {Newell},\ and\ \citenamefont {Castell}}]{silly2006srtio3}%
  \BibitemOpen
  \bibfield  {author} {\bibinfo {author} {\bibfnamefont {F.}~\bibnamefont
  {Silly}}, \bibinfo {author} {\bibfnamefont {D.~T.}\ \bibnamefont {Newell}},\
  and\ \bibinfo {author} {\bibfnamefont {M.~R.}\ \bibnamefont {Castell}},\
  }\bibfield  {title} {\bibinfo {title} {{SrTiO}$_3$(001) reconstructions: the
  (2$\times$2) to \textit{c}(4$\times$4) transition},\ }\href@noop {}
  {\bibfield  {journal} {\bibinfo  {journal} {Surf. Sci.}\ }\textbf {\bibinfo
  {volume} {600}},\ \bibinfo {pages} {219} (\bibinfo {year}
  {2006})}\BibitemShut {NoStop}%
\bibitem [{\citenamefont {Erdman}\ \emph {et~al.}(2002)\citenamefont {Erdman},
  \citenamefont {Poeppelmeier}, \citenamefont {Asta}, \citenamefont
  {Warschkow}, \citenamefont {Ellis},\ and\ \citenamefont
  {Marks}}]{erdman2002structure}%
  \BibitemOpen
  \bibfield  {author} {\bibinfo {author} {\bibfnamefont {N.}~\bibnamefont
  {Erdman}}, \bibinfo {author} {\bibfnamefont {K.~R.}\ \bibnamefont
  {Poeppelmeier}}, \bibinfo {author} {\bibfnamefont {M.}~\bibnamefont {Asta}},
  \bibinfo {author} {\bibfnamefont {O.}~\bibnamefont {Warschkow}}, \bibinfo
  {author} {\bibfnamefont {D.~E.}\ \bibnamefont {Ellis}},\ and\ \bibinfo
  {author} {\bibfnamefont {L.~D.}\ \bibnamefont {Marks}},\ }\bibfield  {title}
  {\bibinfo {title} {The structure and chemistry of the {TiO}$_2$-rich surface
  of {SrTiO}$_3$(001)},\ }\href@noop {} {\bibfield  {journal} {\bibinfo
  {journal} {Nature}\ }\textbf {\bibinfo {volume} {419}},\ \bibinfo {pages}
  {55} (\bibinfo {year} {2002})}\BibitemShut {NoStop}%
\bibitem [{\citenamefont {Herger}\ \emph {et~al.}(2007)\citenamefont {Herger},
  \citenamefont {Willmott}, \citenamefont {Bunk}, \citenamefont
  {Schlep{\"u}tz}, \citenamefont {Patterson},\ and\ \citenamefont
  {Delley}}]{herger2007surfacePRL}%
  \BibitemOpen
  \bibfield  {author} {\bibinfo {author} {\bibfnamefont {R.}~\bibnamefont
  {Herger}}, \bibinfo {author} {\bibfnamefont {P.~R.}\ \bibnamefont
  {Willmott}}, \bibinfo {author} {\bibfnamefont {O.}~\bibnamefont {Bunk}},
  \bibinfo {author} {\bibfnamefont {C.~M.}\ \bibnamefont {Schlep{\"u}tz}},
  \bibinfo {author} {\bibfnamefont {B.~D.}\ \bibnamefont {Patterson}},\ and\
  \bibinfo {author} {\bibfnamefont {B.}~\bibnamefont {Delley}},\ }\bibfield
  {title} {\bibinfo {title} {Surface of strontium titanate},\ }\href@noop {}
  {\bibfield  {journal} {\bibinfo  {journal} {Phys. Rev. Lett.}\ }\textbf
  {\bibinfo {volume} {98}},\ \bibinfo {pages} {076102} (\bibinfo {year}
  {2007})}\BibitemShut {NoStop}%
\bibitem [{\citenamefont {Gerhold}\ \emph {et~al.}(2014)\citenamefont
  {Gerhold}, \citenamefont {Wang}, \citenamefont {Schmid},\ and\ \citenamefont
  {Diebold}}]{Gerhold2014}%
  \BibitemOpen
  \bibfield  {author} {\bibinfo {author} {\bibfnamefont {S.}~\bibnamefont
  {Gerhold}}, \bibinfo {author} {\bibfnamefont {Z.}~\bibnamefont {Wang}},
  \bibinfo {author} {\bibfnamefont {M.}~\bibnamefont {Schmid}},\ and\ \bibinfo
  {author} {\bibfnamefont {U.}~\bibnamefont {Diebold}},\ }\bibfield  {title}
  {\bibinfo {title} {Stoichiometry-driven switching between surface
  reconstructions on {SrTiO}$_3$(001)},\ }\href@noop {} {\bibfield  {journal}
  {\bibinfo  {journal} {Surf. Sci.}\ }\textbf {\bibinfo {volume} {621}},\
  \bibinfo {pages} {L1} (\bibinfo {year} {2014})}\BibitemShut {NoStop}%
\bibitem [{\citenamefont {Goniakowski}\ \emph {et~al.}(2008)\citenamefont
  {Goniakowski}, \citenamefont {Finocchi},\ and\ \citenamefont
  {Noguera}}]{Goniakowski2008}%
  \BibitemOpen
  \bibfield  {author} {\bibinfo {author} {\bibfnamefont {J.}~\bibnamefont
  {Goniakowski}}, \bibinfo {author} {\bibfnamefont {F.}~\bibnamefont
  {Finocchi}},\ and\ \bibinfo {author} {\bibfnamefont {C.}~\bibnamefont
  {Noguera}},\ }\bibfield  {title} {\bibinfo {title} {Polarity of oxide
  surfaces and nanostructures},\ }\href@noop {} {\bibfield  {journal} {\bibinfo
   {journal} {Rep. Prog. Phys.}\ }\textbf {\bibinfo {volume} {71}},\ \bibinfo
  {pages} {016501} (\bibinfo {year} {2008})}\BibitemShut {NoStop}%
\bibitem [{\citenamefont {Koster}\ \emph
  {et~al.}(1998{\natexlab{a}})\citenamefont {Koster}, \citenamefont {Kropman},
  \citenamefont {Rijnders}, \citenamefont {Blank},\ and\ \citenamefont
  {Rogalla}}]{koster1998influence}%
  \BibitemOpen
  \bibfield  {author} {\bibinfo {author} {\bibfnamefont {G.}~\bibnamefont
  {Koster}}, \bibinfo {author} {\bibfnamefont {B.~L.}\ \bibnamefont {Kropman}},
  \bibinfo {author} {\bibfnamefont {G.~J. H.~M.}\ \bibnamefont {Rijnders}},
  \bibinfo {author} {\bibfnamefont {D.~H.~A.}\ \bibnamefont {Blank}},\ and\
  \bibinfo {author} {\bibfnamefont {H.}~\bibnamefont {Rogalla}},\ }\bibfield
  {title} {\bibinfo {title} {Influence of the surface treatment on the
  homoepitaxial growth of {SrTiO}$_3$},\ }\href@noop {} {\bibfield  {journal}
  {\bibinfo  {journal} {Mater. Sci. Eng. B}\ }\textbf {\bibinfo {volume}
  {56}},\ \bibinfo {pages} {209} (\bibinfo {year}
  {1998}{\natexlab{a}})}\BibitemShut {NoStop}%
\bibitem [{\citenamefont {Koster}\ \emph
  {et~al.}(1998{\natexlab{b}})\citenamefont {Koster}, \citenamefont {Kropman},
  \citenamefont {Rijnders}, \citenamefont {Blank},\ and\ \citenamefont
  {Rogalla}}]{koster1998quasi}%
  \BibitemOpen
  \bibfield  {author} {\bibinfo {author} {\bibfnamefont {G.}~\bibnamefont
  {Koster}}, \bibinfo {author} {\bibfnamefont {B.~L.}\ \bibnamefont {Kropman}},
  \bibinfo {author} {\bibfnamefont {G.~J. H.~M.}\ \bibnamefont {Rijnders}},
  \bibinfo {author} {\bibfnamefont {D.~H.~A.}\ \bibnamefont {Blank}},\ and\
  \bibinfo {author} {\bibfnamefont {H.}~\bibnamefont {Rogalla}},\ }\bibfield
  {title} {\bibinfo {title} {Quasi-ideal strontium titanate crystal surfaces
  through formation of strontium hydroxide},\ }\href@noop {} {\bibfield
  {journal} {\bibinfo  {journal} {Appl. Phys. Lett.}\ }\textbf {\bibinfo
  {volume} {73}},\ \bibinfo {pages} {2920} (\bibinfo {year}
  {1998}{\natexlab{b}})}\BibitemShut {NoStop}%
\bibitem [{\citenamefont {Baniecki}\ \emph {et~al.}(2008)\citenamefont
  {Baniecki}, \citenamefont {Ishii}, \citenamefont {Kurihara}, \citenamefont
  {Yamanaka}, \citenamefont {Yano}, \citenamefont {Shinozaki}, \citenamefont
  {Imada}, \citenamefont {Nozaki},\ and\ \citenamefont
  {Kin}}]{baniecki2008photoemission}%
  \BibitemOpen
  \bibfield  {author} {\bibinfo {author} {\bibfnamefont {J.~D.}\ \bibnamefont
  {Baniecki}}, \bibinfo {author} {\bibfnamefont {M.}~\bibnamefont {Ishii}},
  \bibinfo {author} {\bibfnamefont {K.}~\bibnamefont {Kurihara}}, \bibinfo
  {author} {\bibfnamefont {K.}~\bibnamefont {Yamanaka}}, \bibinfo {author}
  {\bibfnamefont {T.}~\bibnamefont {Yano}}, \bibinfo {author} {\bibfnamefont
  {K.}~\bibnamefont {Shinozaki}}, \bibinfo {author} {\bibfnamefont
  {T.}~\bibnamefont {Imada}}, \bibinfo {author} {\bibfnamefont
  {K.}~\bibnamefont {Nozaki}},\ and\ \bibinfo {author} {\bibfnamefont
  {N.}~\bibnamefont {Kin}},\ }\bibfield  {title} {\bibinfo {title}
  {Photoemission and quantum chemical study of {SrTiO}$_3$(001) surfaces and
  their interaction with {CO}$_2$},\ }\href@noop {} {\bibfield  {journal}
  {\bibinfo  {journal} {Phys. Rev. B}\ }\textbf {\bibinfo {volume} {78}},\
  \bibinfo {pages} {195415} (\bibinfo {year} {2008})}\BibitemShut {NoStop}%
\bibitem [{\citenamefont {Sokolovi{\'c}}\ \emph {et~al.}(2024)\citenamefont
  {Sokolovi{\'c}}, \citenamefont {Guedes}, \citenamefont {van Waas},
  \citenamefont {Polley}, \citenamefont {Schmid}, \citenamefont {Diebold},
  \citenamefont {Radovi{\'c}}, \citenamefont {Setv{\'\i}n},\ and\ \citenamefont
  {Dil}}]{sokolovic2024duality}%
  \BibitemOpen
  \bibfield  {author} {\bibinfo {author} {\bibfnamefont {I.}~\bibnamefont
  {Sokolovi{\'c}}}, \bibinfo {author} {\bibfnamefont {E.~B.}\ \bibnamefont
  {Guedes}}, \bibinfo {author} {\bibfnamefont {T.~P.}\ \bibnamefont {van
  Waas}}, \bibinfo {author} {\bibfnamefont {C.}~\bibnamefont {Polley}},
  \bibinfo {author} {\bibfnamefont {M.}~\bibnamefont {Schmid}}, \bibinfo
  {author} {\bibfnamefont {U.}~\bibnamefont {Diebold}}, \bibinfo {author}
  {\bibfnamefont {M.}~\bibnamefont {Radovi{\'c}}}, \bibinfo {author}
  {\bibfnamefont {M.}~\bibnamefont {Setv{\'\i}n}},\ and\ \bibinfo {author}
  {\bibfnamefont {J.}~\bibnamefont {Dil}},\ }\bibfield  {title} {\bibinfo
  {title} {Duality and degeneracy lifting in two-dimensional electron liquids
  on {SrTiO}$_3$(001)},\ }\href@noop {} {\bibfield  {journal} {\bibinfo
  {journal} {arXiv preprint arXiv:2405.18946}\ } (\bibinfo {year}
  {2024})}\BibitemShut {NoStop}%
\bibitem [{\citenamefont {Cohen}(1992)}]{cohen1992origin}%
  \BibitemOpen
  \bibfield  {author} {\bibinfo {author} {\bibfnamefont {R.~E.}\ \bibnamefont
  {Cohen}},\ }\bibfield  {title} {\bibinfo {title} {Origin of ferroelectricity
  in perovskite oxides},\ }\href@noop {} {\bibfield  {journal} {\bibinfo
  {journal} {Nature}\ }\textbf {\bibinfo {volume} {358}},\ \bibinfo {pages}
  {136} (\bibinfo {year} {1992})}\BibitemShut {NoStop}%
\bibitem [{\citenamefont {Resta}\ \emph {et~al.}(1993)\citenamefont {Resta},
  \citenamefont {Posternak},\ and\ \citenamefont
  {Baldereschi}}]{resta1993towards}%
  \BibitemOpen
  \bibfield  {author} {\bibinfo {author} {\bibfnamefont {R.}~\bibnamefont
  {Resta}}, \bibinfo {author} {\bibfnamefont {M.}~\bibnamefont {Posternak}},\
  and\ \bibinfo {author} {\bibfnamefont {A.}~\bibnamefont {Baldereschi}},\
  }\bibfield  {title} {\bibinfo {title} {Towards a quantum theory of
  polarization in ferroelectrics: The case of {KNbO}$_3$},\ }\href@noop {}
  {\bibfield  {journal} {\bibinfo  {journal} {Phys. Rev. Lett.}\ }\textbf
  {\bibinfo {volume} {70}},\ \bibinfo {pages} {1010} (\bibinfo {year}
  {1993})}\BibitemShut {NoStop}%
\bibitem [{\citenamefont {Bousquet}\ \emph {et~al.}(2008)\citenamefont
  {Bousquet}, \citenamefont {Dawber}, \citenamefont {Stucki}, \citenamefont
  {Lichtensteiger}, \citenamefont {Hermet}, \citenamefont {Gariglio},
  \citenamefont {Triscone},\ and\ \citenamefont
  {Ghosez}}]{bousquet2008improper}%
  \BibitemOpen
  \bibfield  {author} {\bibinfo {author} {\bibfnamefont {E.}~\bibnamefont
  {Bousquet}}, \bibinfo {author} {\bibfnamefont {M.}~\bibnamefont {Dawber}},
  \bibinfo {author} {\bibfnamefont {N.}~\bibnamefont {Stucki}}, \bibinfo
  {author} {\bibfnamefont {C.}~\bibnamefont {Lichtensteiger}}, \bibinfo
  {author} {\bibfnamefont {P.}~\bibnamefont {Hermet}}, \bibinfo {author}
  {\bibfnamefont {S.}~\bibnamefont {Gariglio}}, \bibinfo {author}
  {\bibfnamefont {J.-M.}\ \bibnamefont {Triscone}},\ and\ \bibinfo {author}
  {\bibfnamefont {P.}~\bibnamefont {Ghosez}},\ }\bibfield  {title} {\bibinfo
  {title} {Improper ferroelectricity in perovskite oxide artificial
  superlattices},\ }\href@noop {} {\bibfield  {journal} {\bibinfo  {journal}
  {Nature}\ }\textbf {\bibinfo {volume} {452}},\ \bibinfo {pages} {732}
  (\bibinfo {year} {2008})}\BibitemShut {NoStop}%
\bibitem [{\citenamefont {Hwang}\ \emph {et~al.}(2019)\citenamefont {Hwang},
  \citenamefont {Feng}, \citenamefont {Charles}, \citenamefont {Wang},
  \citenamefont {Lee}, \citenamefont {Stoerzinger}, \citenamefont {Muy},
  \citenamefont {Rao}, \citenamefont {Lee}, \citenamefont {Jacobs} \emph
  {et~al.}}]{hwang2019tuning}%
  \BibitemOpen
  \bibfield  {author} {\bibinfo {author} {\bibfnamefont {J.}~\bibnamefont
  {Hwang}}, \bibinfo {author} {\bibfnamefont {Z.}~\bibnamefont {Feng}},
  \bibinfo {author} {\bibfnamefont {N.}~\bibnamefont {Charles}}, \bibinfo
  {author} {\bibfnamefont {X.~R.}\ \bibnamefont {Wang}}, \bibinfo {author}
  {\bibfnamefont {D.}~\bibnamefont {Lee}}, \bibinfo {author} {\bibfnamefont
  {K.~A.}\ \bibnamefont {Stoerzinger}}, \bibinfo {author} {\bibfnamefont
  {S.}~\bibnamefont {Muy}}, \bibinfo {author} {\bibfnamefont {R.~R.}\
  \bibnamefont {Rao}}, \bibinfo {author} {\bibfnamefont {D.}~\bibnamefont
  {Lee}}, \bibinfo {author} {\bibfnamefont {R.}~\bibnamefont {Jacobs}}, \emph
  {et~al.},\ }\bibfield  {title} {\bibinfo {title} {Tuning perovskite oxides by
  strain: Electronic structure, properties, and functions in (electro)
  catalysis and ferroelectricity},\ }\href@noop {} {\bibfield  {journal}
  {\bibinfo  {journal} {Mater. Today}\ }\textbf {\bibinfo {volume} {31}},\
  \bibinfo {pages} {100} (\bibinfo {year} {2019})}\BibitemShut {NoStop}%
\bibitem [{\citenamefont {Uwe}\ \emph {et~al.}(1973)\citenamefont {Uwe},
  \citenamefont {Unoki}, \citenamefont {Fujii},\ and\ \citenamefont
  {Sakudo}}]{uwe1973stress}%
  \BibitemOpen
  \bibfield  {author} {\bibinfo {author} {\bibfnamefont {H.}~\bibnamefont
  {Uwe}}, \bibinfo {author} {\bibfnamefont {H.}~\bibnamefont {Unoki}}, \bibinfo
  {author} {\bibfnamefont {Y.}~\bibnamefont {Fujii}},\ and\ \bibinfo {author}
  {\bibfnamefont {T.}~\bibnamefont {Sakudo}},\ }\bibfield  {title} {\bibinfo
  {title} {Stress induced ferroelectricity in {K}{T}a{O}$_3$},\ }\href@noop {}
  {\bibfield  {journal} {\bibinfo  {journal} {Solid State Commun.}\ }\textbf
  {\bibinfo {volume} {13}},\ \bibinfo {pages} {737} (\bibinfo {year}
  {1973})}\BibitemShut {NoStop}%
\bibitem [{\citenamefont {Vendik}\ \emph {et~al.}(1998)\citenamefont {Vendik},
  \citenamefont {Ter-Martirosyan},\ and\ \citenamefont
  {Zubko}}]{vendik1998microwave}%
  \BibitemOpen
  \bibfield  {author} {\bibinfo {author} {\bibfnamefont {O.~G.}\ \bibnamefont
  {Vendik}}, \bibinfo {author} {\bibfnamefont {L.~T.}\ \bibnamefont
  {Ter-Martirosyan}},\ and\ \bibinfo {author} {\bibfnamefont {S.~P.}\
  \bibnamefont {Zubko}},\ }\bibfield  {title} {\bibinfo {title} {Microwave
  losses in incipient ferroelectrics as functions of the temperature and the
  biasing field},\ }\href@noop {} {\bibfield  {journal} {\bibinfo  {journal}
  {J. Appl. Phys.}\ }\textbf {\bibinfo {volume} {84}},\ \bibinfo {pages} {993}
  (\bibinfo {year} {1998})}\BibitemShut {NoStop}%
\bibitem [{\citenamefont {Setvín}\ \emph {et~al.}(2018)\citenamefont
  {Setvín}, \citenamefont {Reticcioli}, \citenamefont {Poelzleitner},
  \citenamefont {Hulva}, \citenamefont {Schmid}, \citenamefont {Boatner},
  \citenamefont {Franchini},\ and\ \citenamefont
  {Diebold}}]{setvin2018polarity}%
  \BibitemOpen
  \bibfield  {author} {\bibinfo {author} {\bibfnamefont {M.}~\bibnamefont
  {Setvín}}, \bibinfo {author} {\bibfnamefont {M.}~\bibnamefont {Reticcioli}},
  \bibinfo {author} {\bibfnamefont {F.}~\bibnamefont {Poelzleitner}}, \bibinfo
  {author} {\bibfnamefont {J.}~\bibnamefont {Hulva}}, \bibinfo {author}
  {\bibfnamefont {M.}~\bibnamefont {Schmid}}, \bibinfo {author} {\bibfnamefont
  {L.~A.}\ \bibnamefont {Boatner}}, \bibinfo {author} {\bibfnamefont
  {C.}~\bibnamefont {Franchini}},\ and\ \bibinfo {author} {\bibfnamefont
  {U.}~\bibnamefont {Diebold}},\ }\bibfield  {title} {\bibinfo {title}
  {Polarity compensation mechanisms on the perovskite surface
  {K}{T}a{O}$_3$(001)},\ }\href@noop {} {\bibfield  {journal} {\bibinfo
  {journal} {Science}\ }\textbf {\bibinfo {volume} {359}},\ \bibinfo {pages}
  {572} (\bibinfo {year} {2018})}\BibitemShut {NoStop}%
\end{thebibliography}%
\pagebreak{}

\end{document}